\documentclass[
aps,pra,
twocolumn
]{revtex4}
\usepackage{graphicx}
\usepackage{subfigure}
\usepackage{amsmath}
\usepackage{amssymb}
\usepackage{amsthm}
\usepackage{braket}
\usepackage{color}
\usepackage{microtype}
\usepackage{tikz}
\usepackage{makecell}
\usepackage{hyperref}
\hypersetup{colorlinks,linkcolor={blue},citecolor={red},urlcolor={blue}}
\usepackage{enumitem}
\usepackage{float}
\usepackage{array}
\usepackage{booktabs}

\theoremstyle{plain}

\theoremstyle{plain}

\begin{document}

\title{Correcting biased noise using Gottesman-Kitaev-Preskill repetition code with noisy ancilla}

\author{Zhifei Li}
\author{Daiqin Su}
\email{sudaiqin@hust.edu.cn}

\affiliation{ MOE Key Laboratory of Fundamental Physical Quantities Measurement, Hubei Key Laboratory of Gravitation and Quantum Physics, PGMF, Institute for Quantum Science and Engineering, School of Physics, Huazhong University of Science and Technology, Wuhan 430074, China}

\date{\today}

\begin{abstract}
Concatenation of a bosonic code with a qubit code is one of the promising ways to achieve fault-tolerant quantum computation. As one of the most important bosonic codes, Gottesman-Kitaev-Preskill (GKP) code is proposed to correct small displacement error in phase space. If the noise in phase space is biased, square-lattice GKP code can be concatenated with XZZX surface code or repetition code that promises a high fault-tolerant threshold to suppress the logical error. In this work, we study the performance of GKP repetition codes with physical ancillary GKP qubits in correcting biased noise. We find that there exists a critical value of noise variance for the ancillary GKP qubit such that the logical Pauli error rate decreases when increasing the code size. Furthermore, one round of GKP error correction has to be performed before concatenating with repetition code. Our study paves the way for practical implementation of error correction by concatenating GKP code with low-level qubit codes. 
\end{abstract}

\maketitle

\section{Introduction}

Noise is the main hindrance to achieve large-scale fault-tolerant quantum computation. Quantum error correcting code is introduced to correct errors by using redundancy in the Hilbert space~\cite{PhysRevA.52.R2493,gottesman1997stabilizer,terhal2015quantum}. Bosonic codes protect finite-dimensional logical space by encoding it in an infinite-dimensional bosonic quantum system~\cite{weedbrook2012gaussian,braunstein2005quantum}, e.g., a simple harmonic oscillator. Compared to the standard qubit codes that encode a single logical qubit using many physical qubits, the bosonic code is more hardware efficient and is subject to a smaller number of noisy channels~\cite{PhysRevA.97.032346,cai2021bosonic}. Currently well established bosonic codes include Gottesman-Kitaev-Preskill (GKP) code~\cite{PhysRevA.64.012310,grimsmo2021quantum}, cat code~\cite{PhysRevA.59.2631,PhysRevA.68.042319}, binomial code~\cite{PhysRevX.6.031006,hu2019quantum,albert2018performance}, and rotation-symmetric code~\cite{PhysRevX.10.011058,endo2022quantum}. The GKP code is one of the most promising bosonic codes, which corrects small displacement errors in phase space and also photon loss~\cite{PhysRevA.97.032346,noh2018quantum}. Although the GKP code has been proposed for two decades~\cite{PhysRevA.64.012310}, it is prepared only recently in ion-trapped~\cite{fluhmann2019encoding, de2022error} and superconducting~\cite{campagne2020quantum} platforms, and is used to extend the decoherence time of the logical qubit through error correction. The GKP code has promising advantages in optical quantum information processing~\cite{PhysRevLett.123.200502}, however optical GKP states have not been experimentally generated due to the stringent requirement for strong nonlinearity, though various preparation schemes have been proposed~\cite{vasconcelos2010all, PhysRevA.100.052301, eaton2019non, PhysRevLett.128.170503, PhysRevX.13.031001}.  

In order to achieve fault tolerance, the common strategy is to concatenate the GKP code with qubit codes to further suppress the logical error. Examples include concatenation with surface/toric codes~\cite{PhysRevX.8.021054, PhysRevA.99.032344, PhysRevA.102.052408, noh2020fault, Bourassa2021blueprintscalable, PRXQuantum.3.010315}, color code~\cite{bombin2006topological,fowler2011two,PhysRevA.104.062434} etc. Concatenation with qubit codes with a high threshold enables a low squeezing threshold for the GKP states around 10 dB~\cite{PhysRevX.8.021054}, which is within the reach of near-term technologies. A variant of the original surface code, known as the XZZX surface code~\cite{bonilla2021xzzx}, has recently been shown to have a higher threshold for biased noise. It is expected that concatenation of GKP code with XZZX surface code would enable lower squeezing threshold if the displacement error is biased~\cite{zhang2023xzzx}. This can happen in two cases, either the noise is biased and a square-lattice (isotropic) GKP code is used, or the noise is isotropic and a biased GKP code is used. However, syndrome measurement and decoding are still complicated for the XZZX surface code~\cite{higgott2022pymatching}, which therefore consume more physical and computational resources. A relatively easier scheme to suppress biased noise is to concatenate GKP code with repetition code~\cite{stafford2022biased}, which requires easier syndrome measurement and decoding, and has a higher threshold. In Ref.~\cite{stafford2022biased}, the error threshold has been estimated for biased GKP repetition code with isotropic noise, which outperforms the biased planar surface code~\cite{PhysRevA.102.052408}. However, both the data and ancillary GKP qubits are assumed to be ideal, namely, with infinitely energy. The error threshold as derived in Ref.~\cite{stafford2022biased} therefore only provides an upper bound, and the requirement is more stringent when the imperfections from the ancillary GKP qubits are taken into account. 

In this work, we study the concatenation of square-lattice GKP code with classical repetition code to correct biased displacement errors, where both the data and ancillary GKP qubits are physical. The error correction procedure consists of four steps: encoding, one round GKP error correction, syndrome measurement on repetition code and recovery operation according to the measurement outcomes. We find that the GKP error correction before concatenation in general increases the error rate of the GKP code and modifies the error profile, however, it is necessary in order to exploit the power of code concatenation. We also find that the logical Pauli error rate decreases when increasing the size of the repetition code if the noise variance of the ancillary GKP qubits is sufficiently small, while the the logical Pauli error rate increases when increasing the size of the repetition code if the noise variance is too large. This implies that there exists a critical value of noise variance for the ancillary GKP qubits below which the code concatenation shows advantages. Our results set an upper bound for the noise variance of the ancillary GKP qubit such that the concatenation with repetition code is useful. 

The paper is organized as follows. In Sec.~\ref{sec:2} we briefly review the ideal and physical GKP states, and introduce the biased noise model. We then discuss the GKP error correction with physical ancillary GKP qubits to correct small displacement error in position space in Sec.~\ref{sec:3}. In Sec.~\ref{sec:4} we concatenate the GKP code with repetition code to reduce the logical Pauli error rate in position space and estimate the critical value of noise variance for the ancillary GKP qubit. Then in Sec.~\ref{sec:5} we use GKP repetition code to correct biased noise to reduce the overall logical error rate, taking into account the displacement errors in both position space and momentum space. We finally conclude in Sec.~\ref{sec:6}. 

\section{Background}\label{sec:2}

\subsection{Preliminaries}\label{sec:2A}

We briefly review the notations used throughout this paper~\cite{gerry2005introductory,weedbrook2012gaussian}. The annihilation and creation operators of a single bosonic mode are denoted as $\hat{a}$ and $\hat{a}^\dag$, respectively, and satisfy the commutation relation $[\hat{a}, \hat{a}^\dag] = 1$. The position quadrature $\hat{q}$ and momentum quadrature $\hat{p}$ are defined as
\begin{eqnarray}
\hat{q} = \frac{1}{\sqrt{2}} (\hat{a} + \hat{a}^\dag), ~~~~~~ \hat{p} = \frac{i}{\sqrt{2}} (\hat{a}^\dag - \hat{a}), 
\end{eqnarray}
and they satisfy the commutation relation $[\hat{q}, \hat{p}] = i$, where we have used the unit $\hbar = 1$. 
This definition implies that the variance of the vacuum state is normalized to $1/2$.

The displacement operator $\hat{D}(\alpha) = e^{\alpha \hat{a}^\dag - \alpha^* \hat{a}}$, with $\alpha$ a complex number, represents a displacement of the quantum state in phase space. It is also useful to define $\hat{X}(u)$ and $\hat{Z}(v)$ as
\begin{eqnarray}
\hat{X}(u) = e^{-i u \hat{p}}, ~~~~~~ \hat{Z}(v) = e^{i v \hat{q}}, 
\end{eqnarray}
which represent a displacement of the position quadrature with $u$ and a displacement of the momentum quadrature with $v$, namely, 
\begin{eqnarray}
\hat{X}^\dag (u) \hat{q} \hat{X} (u) = \hat{q} + u, ~~~~~~ \hat{Z}^\dag (v) \hat{p} \hat{Z} (v) = \hat{p} + v.  
\end{eqnarray}
If $\ket{s}_q$ and $\ket{s}_p$ are the position and momentum eigenstates, respectively, then
\begin{eqnarray}
\hat{X}(u) \ket{s}_q = \ket{s + u}_q, ~~~~~~ \hat{Z}(v) \ket{s}_p = \ket{s + v}_p.  
\end{eqnarray}
A general displacement operator can be written as
\begin{eqnarray}\label{eq:displace}
\hat{D}(u, v) = e^{-i u \hat{p} + i v \hat{q}} = e^{i u v/2} \hat{X}(u) \hat{Z}(v). 
\end{eqnarray}

The squeezing operator $\hat{S}(s)$ is defined as
\begin{eqnarray}
\hat{S}(s) = e^{ i \ln (s) (\hat{q} \hat{p} + \hat{p} \hat{q})/2 }, 
\end{eqnarray}
where $s > 0$ is the squeezing factor, and its action on the position and momentum quadratures is 
\begin{eqnarray}
\hat{S}^\dag (s) \hat{q} \hat{S} (s) = \frac{1}{s} \, \hat{q}, ~~~~~~ \hat{S}^\dag (s) \hat{p} \hat{S} (s) = s \, \hat{p}. 
\end{eqnarray}
It is evident that if $s>1$ then the position is squeezed and the momentum is anti-squeezed, while if $s<1$ then the position is anti-squeezed and the momentum is squeezed. A squeezed vacuum state is obtained by applying the squeezing operator to the vacuum state, 
\begin{eqnarray}
\ket{s}_{\rm sq} = \hat{S}(s) \ket{0}. 
\end{eqnarray}
The variance of any quadrature in the vacuum state is the same, $\sigma_{\rm vac}^2 = 1/2$;
while in the squeezed vacuum state, the variances of the position and momentum are not the same and are given by
\begin{eqnarray}
\sigma_{\rm q}^2 = \frac{1}{2 s}, ~~~~~~ \sigma_{\rm p}^2 = \frac{1}{2} s, 
\end{eqnarray}
respectively. 

A two-mode unitary operator that is frequently used throughout this paper is 
\begin{eqnarray}
\hat{U}_2 = e^{-i \hat{q}_1 \hat{p}_2},
\end{eqnarray}
which is known as the SUM gate~\cite{PhysRevA.64.012310}. The action of the SUM gate on the position and momentum quadratures is given by
\begin{eqnarray}\label{eq:SUM rule}
&&\hat{q}_1 \rightarrow \hat{q}_1, \quad \hat{p}_1 \rightarrow \hat{p}_1 - \hat{p}_2, \nonumber\\
&&\hat{q}_2 \rightarrow \hat{q}_2 + \hat{q}_1, \quad \hat{p}_2 \rightarrow \hat{p}_2.
\end{eqnarray}


The state of a continuous-variable (CV) quantum system can be represented by the Wigner function, which is defined in the $q$-$p$ phase space as
\begin{eqnarray}\label{eq:WignerDf}
W(q, p) = \frac{1}{\pi} \int {\rm d} y \, e^{-2 i p y} {}_q \bra{q - y} \hat{\rho} \ket{q + y}_q,
\end{eqnarray}
where $\hat{\rho}$ is the density matrix. 

\subsection{GKP states}\label{sec:2B}

The ideal GKP states are common eigenstates of two commuting operators, $\hat{S}_q = \hat{X}(2\sqrt{\pi})$ and $\hat{S}_p = \hat{Z}(2\sqrt{\pi})$, known as the stabilizers~\cite{PhysRevA.64.012310}. Therefore, they form a two-dimensional code subspace of an infinite-dimensional Hilbert space of a bosonic system, the computational bases of which can be chosen as 
\begin{eqnarray}
\ket{\bar{j}} = \sum_{n = - \infty}^{+ \infty} \ket{(2 n + j) \sqrt{\pi}}_q,
\end{eqnarray}
where $j = 0, 1$, and the subscript ``$q$" indicates a position eigenstate. An arbitrary ideal GKP state is a linear superposition of $\ket{\bar{0}}$ and $\ket{\bar{1}}$. The wave functions of the basis states $\ket{\bar{0}}$ and $\ket{\bar{1}}$ in position space are given by
\begin{eqnarray}
\bar{\psi}_j (q) = \sum_{n = - \infty}^{+ \infty} \delta \big(q - (2 n + j) \sqrt{\pi} \, \big).
\end{eqnarray}
It is evident that the wave function of an ideal GKP state is a superposition of a sequence of Dirac delta functions with a fixed period $2\sqrt{\pi}$.

Ideal GKP states are unphysical and cannot be prepared in experiments, though sometimes it is easier to analyze them and some instructive results can be derived. A physical GKP state is a linear superposition of squeezed coherent states with finite squeezing. Mathematically, the physical GKP state can be obtained through coherently superposing randomly displaced ideal GKP states that follow a given probability distribution~\cite{PhysRevA.64.012310,grimsmo2021quantum}, namely, 
\begin{equation}\label{eq:physicalGKP}
\ket{\tilde{\psi}} = N \int {\rm d} u {\rm d} v \, \eta (u, v) e^{ - i u \hat{p} + i v \hat{q} } \ket{\bar{\xi}},
\end{equation}
where $\ket{\bar{\xi}}$ is an ideal GKP state, $N$ is a normalization factor and $\eta (u, v)$ is the probability amplitude describing the quantum diffusion process, which is chosen as a two-dimensional Gaussian distribution throughout this paper, 
\begin{equation}\label{eq:eta}
\eta (u, v) = \frac{1}{\sqrt{\pi \kappa \Delta } } \exp \left [ -\frac{1}{2}\left ( \frac{u^{2}}{\Delta^{2}}+\frac{v^{2}}{\kappa^{2}}  \right )  \right ],
\end{equation}
with $\Delta$ and $\kappa$ the standard deviations of the Gaussian distribution. It can be shown that the physical GKP state defined in Eq.~\eqref{eq:physicalGKP} is normalizable (see Appendix~\ref{sec:appA} for details) and therefore contains a finite amount of energy. In the limit of $\Delta \rightarrow 0$ and $\kappa \rightarrow 0$, the physical computational basis states can be approximated as
\begin{eqnarray}
\tilde{\psi}_j (q) = \bigg( \frac{4 \kappa^2}{\pi \Delta^2} \bigg)^{1/4} \sum_{n = - \infty}^{+ \infty} e^{- \pi (2 n + j)^2 \kappa^2} \nonumber\\
\times \exp \bigg\{-\frac{ \big[ q - (2n+j) \sqrt{\pi} \, \big]^2 }{2 \Delta^2} \bigg\}. 
\end{eqnarray}

\subsection{Noise model}\label{sec:2C}

The noise model that we consider is an anisotropic Gaussian displacement channel (GDC), namely, the noise in one quadrature and its conjugate quadrature may not be the same. Since we use the square-lattice GKP code, a biased logical error will be induced due to the anisotropic GDC, which is further corrected by concatenating with repetition code. This is equivalent to the scheme where an isotropic GDC is considered while the GKP code is biased~\cite{stafford2022biased}. 

\begin{figure}
\subfigure[$\Delta=0.25,r=1$] {\includegraphics[width=0.213\textwidth]{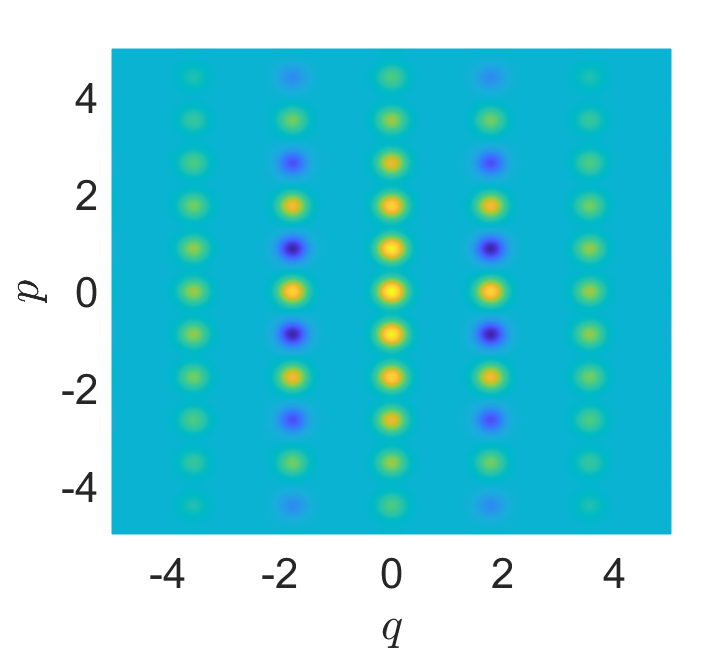}}
\subfigure[$\Delta=0.25,r=\sqrt{2}$] {\includegraphics[width=0.26\textwidth]{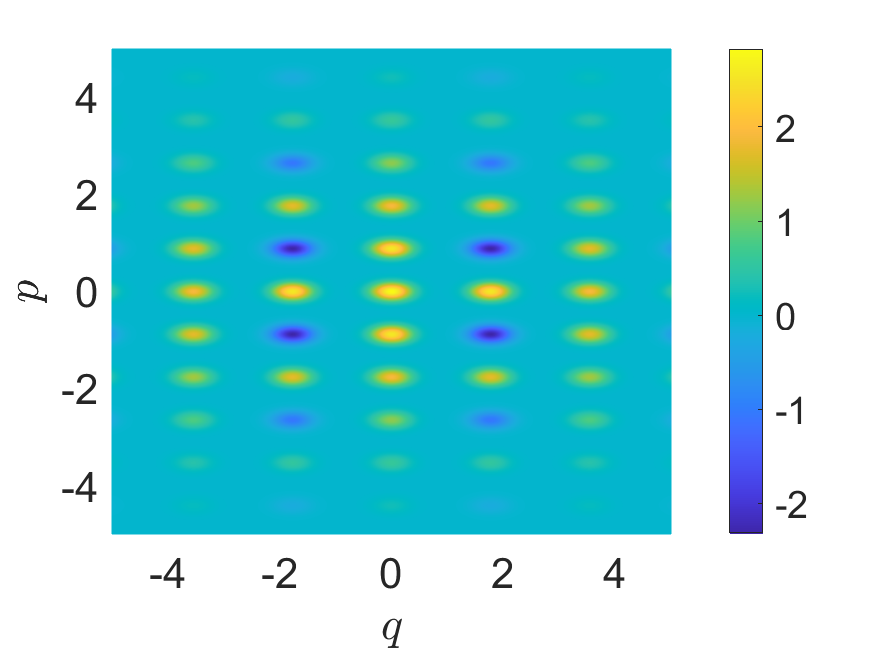}}
\caption{ Wigner function for physical GKP states. (a) Wigner function of GKP state with unbiased noise. (b) Wigner function of GKP state with biased noise. The noise of momentum quadrature is suppressed, while the noise of position quadrature is amplified. }
\label{fig:wigner func}
\end{figure}

Suppose $\hat{\rho}_0$ is an input density matrix, then the output density matrix $\hat{\rho}$ after the anisotropic GDC is
\begin{eqnarray}
\hat{\rho} = \mathcal{N}_f (\hat{\rho}_0) = \int {\rm d} u {\rm d} v f(u, v) \hat{D}(u, v) \hat{\rho}_0 \hat{D}^\dag (u, v),
\end{eqnarray}
where $\hat{D}(u, v)$ is the general displacement operator defined in Eq.~\eqref{eq:displace}, and $f(u, v)$ is a bivariate Gaussian distribution, 
\begin{equation}
f (u,v) = \frac{1}{\pi \delta_q \delta_p } \exp \left ( -\frac{u^{2}}{\delta_q^{2}} - \frac{v^{2}}{\delta_p^{2}}   \right ),
\end{equation}
with $\delta_q$ and $\delta_p$ the standard deviation of the position and momentum quadratures, respectively. According to the definition of the GDC, the input state is imposed a random displacement $\hat{D}(u, v)$ each time, and the output state is an incoherent mixture of all possible displacements. This implies that one can focus on a specific displacement each time when analyzing the error correction.

Under the GDC, the Wigner function is transformed in a simple way,
\begin{eqnarray}
W(q, p; \hat{\rho}) = \int {\rm d} u {\rm d} v f(u - q, v - p) W_0(u, v; \hat{\rho}_0).
\end{eqnarray}
This shows that the output Wigner function is a convolution of the input Wigner function and the error distribution function. Although the physical GKP state is not Gaussian, its Wigner function is a weighted sum of a sequence of Gaussian functions. Since the convolution of two Gaussian functions gives also a Gaussian function, the Wigner function of a physical GKP state after the GDC is still a weighted sum of a sequence of Gaussian functions. In addition, the variances of the output Gaussian functions are the sum of the variances of the input Gaussian functions and error distribution function. Therefore, the Wigner function is blurred after the GDC.  

Note that the GDC is essentially different from the coherent superposition of random displacements that involved in defining physical GKP states in Eq.~\eqref{eq:physicalGKP}, in particular, a physical GKP state cannot be generated by simply passing an ideal GKP state through a GDC (see Appendix~\ref{sec:appA} for details). The physical GKP states have finite energy while the states obtained by passing ideal GKP states through GDC are unphysical and have infinite energy. However, these two sets of state have exactly the same noise property if we set 
\begin{equation}\label{eq:physical fuv}
f (u, v) = \left | \eta(u,v) \right |^2.
\end{equation}
This is clear when comparing their Wigner functions: the Wigner function of the former is a weighted sum of a sequence of Gaussian functions, with the weight decreases exponentially for large integers; the Wigner function of the latter is a sum of the same Gaussian functions but with equal weight for all integers. Therefore, we can treat the noise in the physical GKP state in the same way as we treat the noise from the GDC when the envelope is irrelevant. 

Since the displacement errors in position and momentum spaces are independent, the error distribution of the physical GKP state given by Eq.~\eqref{eq:physical fuv} can be factored into the position and momentum parts,
\begin{equation}\label{eq:physical f}
f (u,v) = 
f_q(u)f_p(v),
\end{equation}
where the error distribution in position and momentum space is respectively given by
\begin{equation}\label{eq:intrinsic error distribution}
f_q(u)=\frac{1}{\sqrt{\pi}\Delta}e^{-\frac{u^{2}}{\Delta^{2}} }, \quad f_p(v)=\frac{1}{\sqrt{\pi}\kappa}e^{-\frac{v^{2}}{\kappa^{2}}}.
\end{equation}
The displacement noise we usually consider is unbiased, i.e., $\Delta=\kappa$. In this paper, we consider GKP code with biased noise, where the noise in one quadrature is suppressed while that in the conjugate quadrature is amplified. The error distribution of biased noise can be parameterized as
\begin{equation}\label{eq:biased intrinsic error distribution}
f_q(u)=\frac{1}{\sqrt{\pi} \, r\Delta}e^{-\frac{u^{2}}{(r\Delta)^{2}} }, \quad f_p(v)=\frac{r}{\sqrt{\pi} \, \Delta}e^{-\frac{v^{2}}{(\Delta/r)^{2}}},
\end{equation}
where $r$ is a real positive number and represents the bias level. By choosing $r>1$, the noise in momentum space is suppressed while that in position space is amplified. In this case, we can concatenate GKP code with repetition code to suppress the logical Pauli error induced by the large displacement error in position space. The Wigner function of GKP code with unbiased and biased noise is shown in Fig.~\ref{fig:wigner func}, where we choose $\Delta=0.25,r=1$ and $\Delta=0.25,r=\sqrt{2}$ for comparison.

\section{GKP error correction with noisy ancillary qubits}\label{sec:3}

In this section, we discuss the correction of displacement error in position space using GKP code with ideal and noisy ancillary GKP qubits. 

\subsection{GKP error correction with ideal ancilla}\label{sec:3A}

\begin{figure}
\includegraphics[width=0.8 \columnwidth]{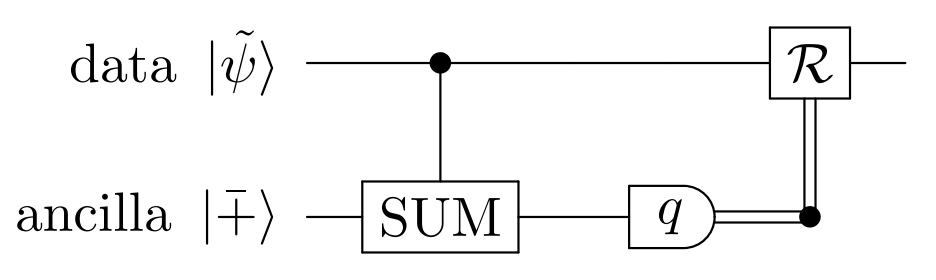}
\caption{Quantum circuit for GKP error correction. The ancillary GKP qubit is prepared in state $\ket{\bar{+}}$ and then couples with the data GKP qubit via a SUM gate. The position shift of the data qubit propagates to the ancillary qubit and is detected by measuring the position of the ancillary qubit. Recovery is finally executed according to the measurement outcome.}
\label{fig:SUM gate}
\end{figure}

The quantum error correction circuit using SUM gate is shown in Fig.~\ref{fig:SUM gate}. The ancillary qubit couples with the data qubit via the SUM gate, then its position quadrature is measured and the measurement outcome is fed forward to the data qubit~\cite{PhysRevA.64.012310}. Suppose the data qubit is prepared in a physical GKP state $\ket{\tilde{\psi}}$ with $\eta(u,v)$ given by Eqs.~\eqref{eq:physicalGKP} and \eqref{eq:eta}. Now we only consider correcting the displacement in position space and rewrite $\ket{\tilde{\psi}}$ as
\begin{equation}
\ket{\tilde{\psi}} = \int {\rm d} v \, \eta (v)e^{iv\hat{q}}\cdot\int {\rm d} u \, \eta (u) e^{-iuv/2}\ket{\psi (u)},
\end{equation}
where $\ket{\psi (u)}$ is an ideal GKP state with position shifted by $u$, 
\begin{equation*}
\left | \psi(u)  \right \rangle=\alpha \sum_{s}^{} \left | 2s\sqrt{\pi}+u   \right \rangle _{q_1}+\beta \sum_{s}^{} \left | (2s+1)\sqrt{\pi}+u   \right \rangle _{q_1}.
\end{equation*}
The ancillary qubit is assumed to be in the ideal GKP $\ket{\bar{+}}$ state,
\begin{equation}
 \ket{\bar{+}} = \sum_{k}^{}\left | k \sqrt{\pi}   \right \rangle _{q_{2}}.
\end{equation}
According to the transformation rule of the SUM gate given by Eq.~\eqref{eq:SUM rule}, the state after the SUM gate is given by
\begin{eqnarray}\label{eq:SUM-state}
&&\int {\rm d} v \, \eta (v)e^{iv\hat{q}}\cdot\int {\rm d} u \, \eta(u)e^{-iuv/2} \nonumber\\
&& \times \bigg[ \alpha \sum_{s,k} \ket{ 2s\sqrt{\pi}+u }_{q_1}\ket{ (2s+k)\sqrt{\pi}+u }_{q_2} \nonumber\\ 
&& + \beta \sum_{s,k} \ket{(2s+1)\sqrt{\pi}+u}_{q_1}\ket{(2s+k+1)\sqrt{\pi}+u}_{q_2} \bigg] \nonumber\\
&&= \int {\rm d} v \, \eta (v)e^{iv\hat{q}}\cdot\int {\rm d} u \, \eta(u) e^{-iuv/2} \nonumber\\
&& \times \ket{\psi (u)} \bigg( \sum_k \ket{ k \sqrt{\pi}+u }_{q_2} \bigg). 
\end{eqnarray}
The homodyne measurement of the ancillary qubit gives a fixed value for $\hat{q}_2$,
\begin{equation}
q_2 =  k \sqrt{\pi} + u,
\end{equation}
with $k$ an integer. This implies that the superposition of different displacements is destroyed and the state in Eq.~\eqref{eq:SUM-state} collapses to a component with a fixed $u$. However, the superposition between GKP states $\ket{\bar{0}}$ and $\ket{\bar{1}}$ (shifted by $u$) is preserved, since they cannot be distinguished by the measurement outcome. 

\begin{figure}
\includegraphics[width=1\columnwidth]{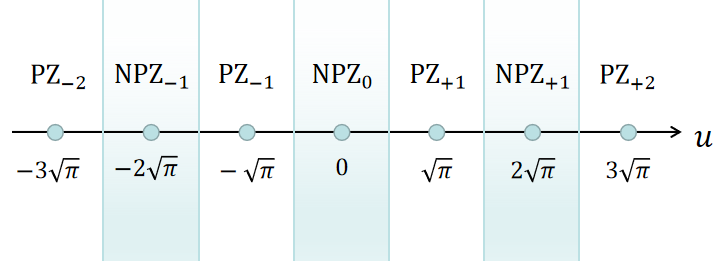}
\caption{Distribution of no Pauli error zone (NPZ) and Pauli error zone (PZ). $\text{PZ}_{0}$ is not defined for the sake of symmetry.}
\label{fig:NPZandPZ}
\end{figure}

Since we consider small displacement errors, so with a high probability $q_2$ deviates from $k \sqrt{\pi}$ in a small amount. Therefore, we infer the true value of $u$ by subtracting from $q_2$ the nearest $k \sqrt{\pi}$. Define a function $g(x)$, which gives the distance between $x$ and its nearest $k \sqrt{\pi}$, 
\begin{equation}
g(x)=x-k\sqrt{\pi}, ~~~\text{ for }(k-\frac{1}{2} )\sqrt{\pi}\le x<  (k+\frac{1}{2} )\sqrt{\pi}. 
\end{equation}
So our guess for the value of $u$ is $g(q_2)$ and we apply a displacement $-g(q_2)$ to the data qubit in order to correct the error. With a high probability the displacement error can be corrected successfully, while sometimes the error correction procedure can introduce a large displacement error and therefore result in a logical Pauli error. Define the residual displacement of the GKP state after the SUM gate and feed forward as
\begin{equation}
u'=u-g(q_2)=u-g(u).
\end{equation}
If $\left | u-2k\sqrt{\pi} \right |<\sqrt{\pi}/2 $, then $g(u)=u-2k\sqrt{\pi}$ and $u'=u-(u-2k\sqrt{\pi})=2k\sqrt{\pi}$, which means a stabilizer is applied to the GKP state and no error occurs. If $\left | u-(2k+1)\sqrt{\pi} \right |<\sqrt{\pi}/2 $, then $g(u)=u-(2k+1)\sqrt{\pi}$ and $u'=u-[u-(2k+1)\sqrt{\pi}]=(2k+1)\sqrt{\pi}$, which means a stabilizer and a logical Pauli operator $\bar{X}$ that flips the computational basis states are applied to the GKP state and a logical Pauli error occurs. We divide the displacement error in position space into two different zones, denoted as Pauli error zone (PZ) and no Pauli error zone (NPZ), according to whether they lead to a logical Pauli error or not, 
\begin{eqnarray}
\text{PZ}_{~} &=& \left \{u: | u-(2k+1)\sqrt{\pi} |<\frac{\sqrt{\pi} }{2}, ~ k \in \mathbb{Z} \right \},  \nonumber\\
\text{NPZ} &=& \left \{u: | u-2k\sqrt{\pi} |<\frac{\sqrt{\pi} }{2}, ~ k \in \mathbb{Z} \right \}.
\end{eqnarray}
For narrative convenience we define a serial number for PZ and NPZ, 
\begin{eqnarray}
\text{PZ}_{m~} &=& \left [ (2m-\frac{m}{\left | m \right | } - \frac{1}{2})\sqrt{\pi} , (2m-\frac{m}{\left | m \right | } + \frac{1}{2}) \sqrt{\pi} \right ), \nonumber\\
\text{NPZ}_{m} &=& \left [ 2m\sqrt{\pi}-\frac{\sqrt{\pi}}{2} , 2m\sqrt{\pi}+\frac{\sqrt{\pi}}{2} \right ), 
\end{eqnarray}
with $m \in \mathbb{Z}$. 
Note that $\text{PZ}_{0}$ is not defined for the sake of symmetry. The location of NPZ$_m$ and PZ$_m$ is shown in Fig.~\ref{fig:NPZandPZ}. 
\begin{figure}
\includegraphics[width=0.85\columnwidth]{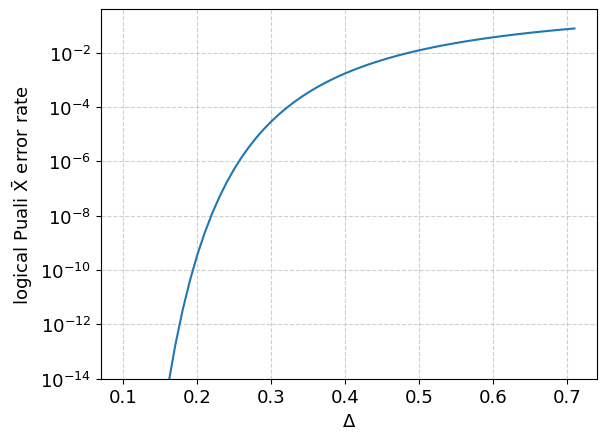}
\caption{Relation between the logical Pauli $\Bar{X}$ error rate and the standard deviation of the probability distribution of the physical GKP state.}
\label{fig:intrinsic error rate}
\end{figure}
With these definitions the error correction procedure can be summarized as follows:
\begin{eqnarray*}
u &\in& \text{NPZ} \Rightarrow u' (\text{mod} \sqrt{\pi} )=0  ~\Rightarrow ~ \text{perfect correction}, \nonumber\\
u &\in& _{~}\text{PZ}_{~} \Rightarrow u' (\text{mod} \sqrt{\pi} )=\sqrt{\pi} ~\Rightarrow~ \text{Pauli}~ \bar{X}~\text{error}.
\end{eqnarray*}
The failure probability of error correction, $P_{\bar{X}}$, which is also known as the logical Pauli $\Bar{X}$ error rate, is the probability that $u$ falls in the PZ.  
\begin{eqnarray}\label{logical error rate X}
P_{\bar{X}} &=& \int_{\text{PZ}} f_q (u) {\rm d} u   = \sum_{n=-\infty }^{+\infty } \int_{\sqrt{\pi}/2+2n\sqrt{\pi} }^{3\sqrt{\pi}/2+2n\sqrt{\pi}} f_q(u) {\rm d} u  \nonumber\\ 
&=& \frac{1}{2} \sum_{n=-\infty }^{+\infty } \left [ \text{erf} \left ( \frac{4n+3}{2 \Delta}  \sqrt{\pi} \right ) - \text{erf} \left ( \frac{4n+1}{2 \Delta}  \sqrt{\pi}  \right )\right ]. \nonumber\\
\end{eqnarray}
The relation between $P_{\bar{X}}$ and $\Delta$ is plotted in Fig.~\ref{fig:intrinsic error rate}. We can see that a smaller $\Delta$, which corresponds to a higher degree of squeezing, leads to a lower logical Pauli $\Bar{X}$ error rate.

\subsection{GKP error correction with physical ancilla}\label{sec:3B}

We now consider a more realistic error correction procedure with physical ancillary GKP qubits. Suppose the variances of the data and ancillary qubit of the position quadrature are $\Delta^2$ and $\tilde{\Delta}^2$, and the displacement errors in position space are $u_1$ and $u_2$, respectively. The probability distribution of $u_1$ and $u_2$ are given by
\begin{equation}\label{eq:distribution u1u2}
f_{q_{1}}(u_1)=\frac{1}{\sqrt{\pi}\Delta }e^{-\frac{u_1^2}{\Delta^2} },\quad  f_{q_{2}}(u_2)=\frac{1}{\sqrt{\pi}\tilde {\Delta} }e^{-\frac{u_2^2}{\tilde {\Delta}^2} }.
\end{equation}
According to the transformation rule of the SUM gate, one can show that the measurement outcome of the ancillary qubit is
\begin{equation}
q_2 =  k \sqrt{\pi} + u_1 + u_2. 
\end{equation}
However, both $u_1$ and $u_2$ are unknown. By using the same procedure as before, we infer the true value of $u_1$ by subtracting from $q_2$ the nearest $k \sqrt{\pi}$. This means our guess for the error in the data qubit is $g(u_1+u_2)$. This is not exactly the same as $u_1$ except that $u_2 = m \sqrt{\pi}$. However, this procedure is acceptable when $u_2$ is sufficiently small. We then apply a displacement $-g(u_1+u_2)$ to the data qubit in order to correct its displacement error. The residual displacement in the data qubit is 
\begin{eqnarray}\label{eq:error distrbution after SUM gate}
u'&=&u_1-g(u_1+u_2) = k\sqrt{\pi} - u_2, \nonumber\\
&& \text{for }(k-\frac{1}{2})\sqrt{\pi}\le u_1+u_2 < (k+\frac{1}{2})\sqrt{\pi}.
\end{eqnarray}
It is evident that the residual displacement $u'$ is continuous, in contrary to the ideal case where $u'$ is discrete. However, one can still define whether a logical Pauli error occurs or not. When $u'$ is close to $2 k \sqrt{\pi}$, no logical Pauli error occurs; when $u'$ is close to $(2 k +1 ) \sqrt{\pi}$, a Pauli $\bar{X}$ error occurs. 

In order to understand the error correcting property with physical ancillary qubit and to evaluate the logical Pauli error rate, one needs to compute the probability distribution of $u'$, which is given by (see Appendix~\ref{sec:appB} for details)
\begin{eqnarray}\label{eq:udot distribution}
F(u') 
&=& \frac{1}{2\sqrt{\pi}\tilde {\Delta} }  \left [ {\rm erf} \left ( \frac{u' + \frac{\sqrt{\pi}}{2}}{\Delta}  \right ) - {\rm erf} \left ( \frac{u' - \frac{\sqrt{\pi}}{2}}{\Delta} \right )  \right ] \nonumber\\
&& \times \sum_{t}^{} {\rm exp} \left [ -\frac{(u' - t \sqrt{\pi})^2}{\tilde{\Delta}^2} \right ]. 
\end{eqnarray}
We can see that $F(u')$ is determined by a modulating term ${\rm erf}[(u'+\sqrt{\pi}/2)/\Delta]- {\rm erf}[(u'-\sqrt{\pi}/2)/\Delta]$ and a wave packet term $\sum_{t}^{}\exp\left [ -\frac{(u' -t\sqrt{\pi})^2}{\tilde{\Delta}^2} \right ] $. The former is determined by the degree of squeezing of the data qubit, while the latter is determined by the degree of squeezing of the ancillary qubit. 

\begin{figure}
\includegraphics[width=0.9\columnwidth]{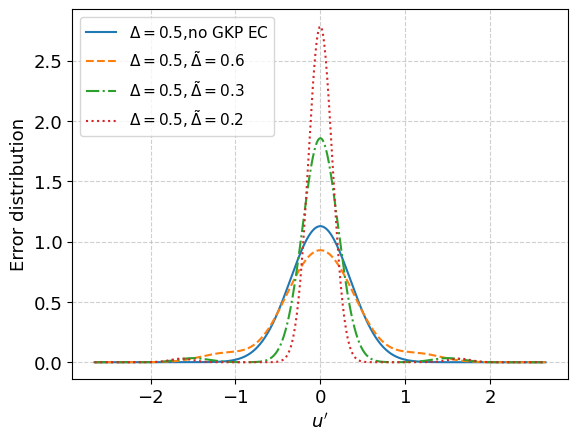}
\caption{Error distribution of GKP state after error correction with a physical ancillary qubit. We choose $\Delta=0.5$ and compare results with several different $\tilde{\Delta}$. The peaks outside $\text{PZ}_{\pm 1}$ are strongly suppressed and a smaller $\tilde{\Delta}$ leads to a narrower peak in $\text{NPZ}_0$, indicating a better correction to the small displacement error.}
\label{fig:SUM distribution}
\end{figure}
\begin{figure}
\includegraphics[width=0.9\columnwidth]{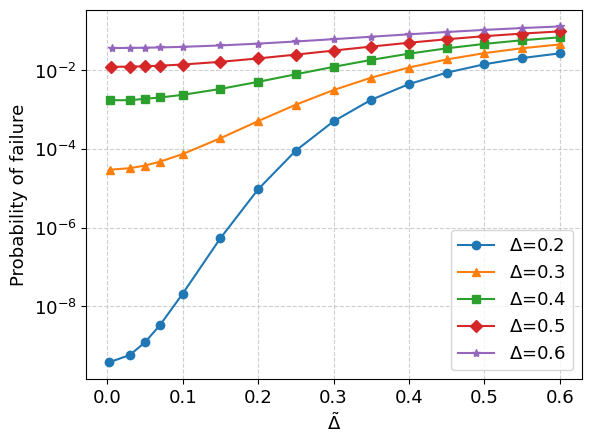}
\caption{Relation between the failure probability $P_{F}(\Delta, \tilde{\Delta})$ with physical ancillary qubit and $\tilde{\Delta}$, with $\Delta$ fixed for each curve. }
\label{fig:SUM failure}
\end{figure}

To have an intuitive feeling of the probability distribution, we plot several examples of $F(u')$ in Fig.~\ref{fig:SUM distribution}. It can be seen that the distribution has a high peak at $u' = 0$ and two low peaks that are located symmetrically with respect to $u' = 0$. The peaks outside the ${\rm PZ}_{\pm 1}$ are strongly suppressed by the modulating term, so the residual displacement outside the ${\rm PZ}_{\pm 1}$ can be neglected. Additionally, a smaller $\tilde{\Delta}$ leads to a narrower distribution of $u'$ in the ${\rm NPZ}_0$, showing a better performance of error correction. This can be understood in an intuitive way: error correction using the SUM gate is basically to substituting the error of the data qubit by the error of the ancillary qubit, hence an ancillary qubit with higher quality naturally leads to a better performance of error correction. If the ancillary qubit is ideal, i.e., $\tilde{\Delta}=0$, then the distribution of $u'$ approaches to a delta function, which means the state can be perfectly corrected.

The error correction is successful when $u'$ is in NPZ and fails when $u'$ is in PZ. Therefore, the failure probability of error correction, namely, the logical Pauli error rate is given by
\begin{eqnarray}
P_{F}(\Delta, \tilde{\Delta}) &=& 
 \sum_{n = -\infty}^{+ \infty} \int_{\sqrt{\pi}/2+2n\sqrt{\pi}}^{3\sqrt{\pi}/2+2n\sqrt{\pi}} F(u') {\rm d} u' 
\nonumber\\
&\approx& 
2\int_{\sqrt{\pi}/2}^{3\sqrt{\pi}/2}F(u') {\rm d} u'. 
\end{eqnarray}
The relation between $P_F(\Delta, \tilde{\Delta})$ and $\tilde{\Delta}$ for a fixed $\Delta$ is plotted in Fig.~\ref{fig:SUM failure}. We can see that $P_F$ monotonically decreases as $\tilde{\Delta}$ decreases, showing that an ancillary GKP qubit with higher quality naturally leads to lower logical Pauli error rate. In addition, it can be shown that
\begin{equation}
P_F(\Delta ,\tilde{\Delta }\to 0 )=P_{\bar{X}}(\Delta). 
\end{equation}
It is an important property that error correction by SUM gate with a physical ancillary qubit always increases logical Pauli error rate as compared to that with an ideal ancillary qubit. In the case of ideal ancillary qubit, a logical Pauli error occurs when $u_1\in {\rm PZ}$ and no error occurs when $u_1\in {\rm NPZ}$. While in the case of physical ancillary qubit, $u_1\in {\rm NPZ}$ may lead to a logical Pauli error because of the presence of an additional displacement $u_2$. Although $u_1\in {\rm PZ}$ may not lead to a logical Pauli error due to the same reason, its probability is much less than the former.

\section{Concatenation with repetition code}\label{sec:4}

In the previous section, we discuss GKP error correction with ideal and physical ancillary GKP qubits, and found that small displacement error can be effectively corrected. The logical Pauli error rate with physical ancillary GKP qubits is generally higher than that with ideal ancillary GKP qubits. To further suppress the logical Pauli error, we concatenate the GKP code with repetition code~\cite{stafford2022biased,fukui2017analog,xu2023qubit}. Note that by concatenating with repetition code we only correct the displacement error in position space.

\subsection{Concatenation with three-qubit repetition code}\label{sec:4A}

\begin{figure}
\includegraphics[width=1.0\columnwidth]{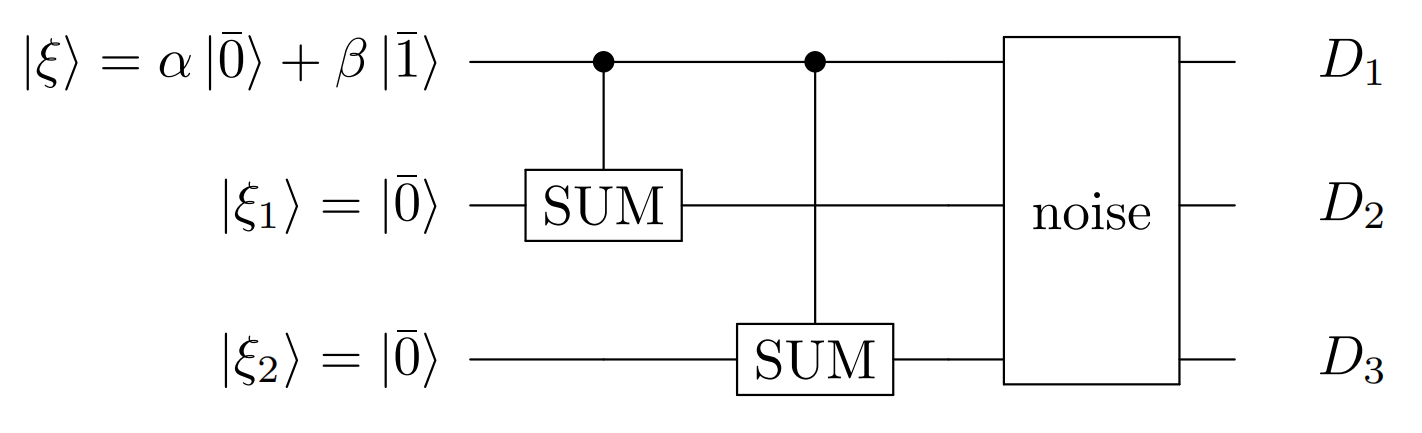}
\caption{Quantum circuit of encoding for 3-qubit GKP repetition code. The three GKP states before encoding are assumed to be ideal. After the encoding, three data qubits entangle with each other. A physical GKP repetition code state is constructed by coherently superposing the ideal GKP states undergoing random displacements.
}
\label{fig:encode 3rep}
\end{figure}
\begin{figure*}
\includegraphics[width=1.6\columnwidth]{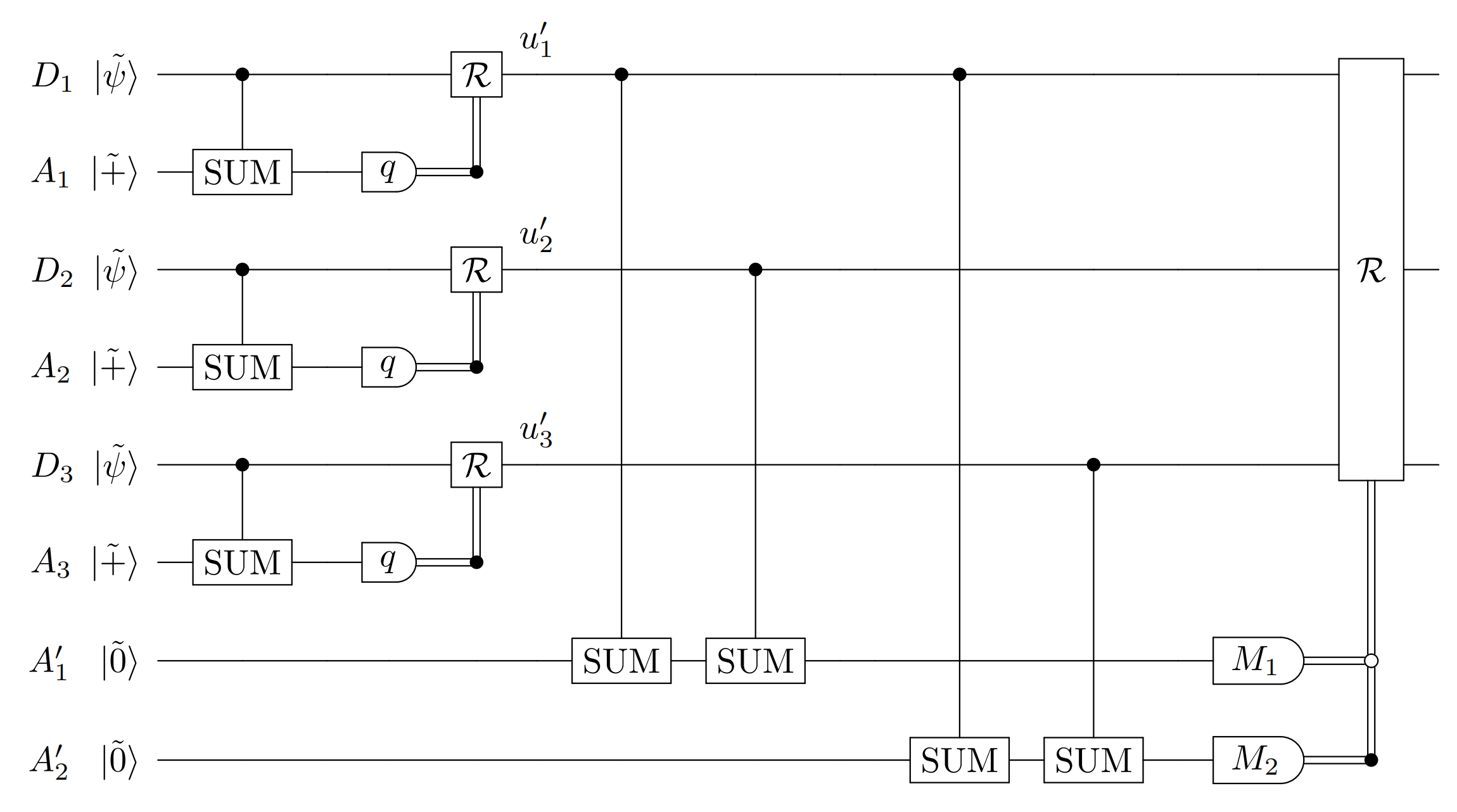}
\caption{Quantum error correction circuit of 3-qubit GKP repetition code. It consists of one round of GKP error correction, syndrome measurement of repetition code, and recovery operation according to the measurement outcomes $M_1$ and $M_2$. Three data qubits $D_1$, $D_2$ and $D_3$ are all physical GKP states with noise variance $\Delta^2$ in position quadrature. Ancillary qubits $A_1$, $A_2$ and $A_3$ are prepared in the state $\ket{\tilde{+}}$, ancillary qubits $A_1^\prime$ and $A_2^\prime$ are prepared in the state $\ket{\tilde{0}}$, and we assume noise variances in position quadrature of all ancillary qubits to be $\tilde{\Delta}^2$. Residual displacements of three data qubits $D_1$, $D_2$ and $D_3$ after the GKP error correction are denoted as $u_1'$, $u_2'$ and $u_3'$, respectively. }
\label{fig:3rep code}
\end{figure*}

In the 3-qubit bit-flip repetition code~\cite{nielsen2001quantum,devitt2013quantum}, three physical qubits are introduced to encode one logical qubit, in particular, the single-qubit state $\alpha \ket{0}+\beta \ket{1}$ is encoded as follows:
\begin{equation}
\ket{\psi}=\alpha \ket{0}+\beta \ket{1} \to \ket{\bar{\psi}} = \alpha \ket{000}+\beta \ket{111}.
\end{equation}
If one of the three qubits was flipped, the flipped qubit can be detected by comparing any two of the three qubits and then applying the majority rule, which is known as the syndrome measurement. Once the flipped qubit is identified, it can be corrected by applying a Pauli $X$ operator. The 3-qubit repetition code can only correct single-qubit bit flip, and bit flip on two or more qubits result in logical error. Denote the bit-flip error rate of a single physical qubit as $p$, then the probability of failure is given by
\begin{equation}
p_{f,\text{3-rep}}^{{\rm class}} =3p^2(1-p)+p^3.
\end{equation}
To realize the comparison between physical qubits without collapsing the encoded state, one needs to introduce ancillary qubits to perform the syndrome measurement. Since one has to identify no error and three single-qubit bit-flip errors, there are $C_3^0+C_3^1=4$ syndromes and therefore two ancillary qubits are needed. The comparison of states of two qubits is implemented by the CNOT gate.

To concatenate the GKP code with repetition code, one needs to replace the standard qubits by the GKP qubits and find a CV gate that corresponds to the CNOT gate. It turns out that the SUM gate we used to perform GKP error correction plays the role as a CNOT gate. The quantum circuit of encoding is shown in Fig.~\ref{fig:encode 3rep}, which is a generalization of the encoding circuit for repetition code. Before encoding, the first GKP qubit is prepared in the state $\ket{\xi} = \alpha \ket{\bar{0}} + \beta \ket{\bar{1}}$, and the other two data qubits are prepared in the same state $\ket{\xi_1} = \ket{\xi_2} = \ket{\bar{0}}$. After the encoding procedure, namely, the application of two SUM gates,  three GKP states become entangled with each other, 
\begin{equation}
\ket{\xi,\xi_1,\xi_2} = (\alpha \ket{\bar{0}} + \beta \ket{\bar{1}}) \ket{\bar{0}}  \ket{\bar{0}}  \to \ket{\bar{\psi}_3}=\alpha\ket{\bar{0}\bar{0}\bar{0}}+\beta\ket{\bar{1}\bar{1}\bar{1}}.
\end{equation}
The state $\ket{\bar{\psi}_3}$ as defined is an ideal GKP repetition code state. One way to construct a physical GKP repetition code state is to coherently superpose the randomly displaced ideal GKP repetition code states, namely, 
\begin{equation}\label{eq:rep-3-code}
\begin{split}
&\ket{\Tilde{\Psi}_3}=\int {\rm d} u_1 {\rm d} v_1 {\rm d} u_2 {\rm d}v_2 {\rm d} u_3 {\rm d} v_3 \, \eta(u_1,v_1)\eta(u_2,v_2 ) \\ 
&\eta(u_3,v_3)e^{i(-u_1\hat{p}_1+v_1\hat{q}_1)}  e^{i(-u_2 \hat{p}_2+v_2 \hat{q}_2)} e^{i(-u_3 \hat{p}_3+v_3 \hat{q}_3)}\ket{\bar{\psi}_3}.
\end{split}
\end{equation}
Here we assume that the displacement in each ideal GKP qubit is independent and follows the same probability distribution. This definition of the physical code state is similar to the definition of a single-qubit physical GKP state~\eqref{eq:physicalGKP}. The GKP repetition code state defined in Eq.~\eqref{eq:rep-3-code} is different from the state generated by applying two SUM gates to three single-qubit physical GKP states, since the latter would generate correlated noise between different GKP qubits. We use the GKP repetition code state~\eqref{eq:rep-3-code} only to seek convenience for calculation, and we will leave the discussion on its experimental preparation for future work.

Error correction is performed after the encoding, which is implemented by the quantum circuit shown in Fig.~\ref{fig:3rep code}. The full process of error correction consists of three steps: one round of GKP error correction, syndrome measurement and feed forward based on the measurement outcome. The three data GKP qubits, denoted as $D_1$, $D_2$, $D_3$, are physical and their noise variances of the position quadrature are the same, which is assumed to be $\Delta^2$. Three ancillary GKP qubits, denoted as $A_1, A_2$ and $A_3$, are introduced to perform the GKP error correction, and they are prepared in the GKP $\ket{\tilde{+}}$ state. Another two ancillary GKP qubits, denoted as $A_1^\prime$ and $A_2^\prime$, are introduced to perform the syndrome measurement, and they are prepared in the GKP $\ket{\tilde{0}}$ state. The noise variances of the position quadrature of all ancillary GKP qubits are assumed to be the same and is $\tilde{\Delta}^2$. The residual displacements of the three data qubits after the GKP error correction are denoted as $u_1'$, $u_2'$, $u_3'$, respectively. Their probability distribution is given by Eq.~\eqref{eq:udot distribution}. 
Denote the displacement errors of the ancillary qubits $A_1^\prime$ and $A_2^\prime$ as $\alpha_1$ and $\alpha_2$, respectively, whose probability distribution is given by
\begin{equation}\label{eq:ancilla distribution}
 f_{q_{i}'}(\alpha _i)=\frac{1}{\sqrt{\pi}\tilde {\Delta} }e^{-\frac{\alpha _i^2}{\tilde {\Delta}^2} }, ~~~~~~i=1,2.
\end{equation}
The purpose of the syndrome measurement is to compare the states of three data GKP qubits, which is implemented by applying four SUM gates that act on the data qubits and the ancillary qubits in an appropriate way, as shown in Fig.~\ref{fig:3rep code}. After these SUM gates, the displacement errors of the ancillary qubits $A_1^\prime$ and $A_2^\prime$ become $u_1'+u_2'+\alpha_1$ and $u_1'+u_3'+\alpha_2$, respectively. Then measurement of the ancillary qubits $A_1^\prime$ and $A_2^\prime$ gives $M_1=2k_1\sqrt{\pi}+u_1'+u_2'+\alpha_1$ and $M_2=2k_2\sqrt{\pi}+u_1'+u_3'+\alpha_2$, with $k_1 \in \mathbb{Z}$ and $k_2 \in \mathbb{Z}$.

\begin{table}
\caption{ Correspondence between syndromes and single-qubit bit-flip errors for the classical 3-qubit repetition code.}
\centering
\begin{tabular}{c|c|c}
\hline \hline
   {\bf syndrome} & {\bf Final state} & {\bf Error} \\
    \hline
   0 0 & ~~$\alpha\ket{000}+\beta\ket{111}$~~ & no error\\
   \hline
   1 1 & ~~$\alpha\ket{100}+\beta\ket{011}$~~ & ~~bit flip on data qubit 1~~ \\
   \hline
   1 0 & $~~\alpha\ket{010}+\beta\ket{101}$~~ & ~~bit flip on data qubit 2~~ \\
   \hline
   0 1 & ~~$\alpha\ket{001}+\beta\ket{110}$~~ & ~~bit flip on data qubit 3~~ \\
\hline
\end{tabular}
\label{tab:classical 3rep error syndrome}
\end{table}

\begin{table}
\caption{Correspondence between syndromes and single-qubit bit-flip errors for the 3-qubit GKP repetition code. The syndromes are defined according to whether $M_1$ and $M_2$ belong to $\text{PZ}$ or $\text{NPZ}$.}
\centering
\renewcommand{\arraystretch}{1.1} 
\begin{tabular}{c|c}
\hline \hline
   {\bf Measurement outcome} & {\bf Error } \\
    \hline
   ~~$M_1\in \text{NPZ}, M_2\in \text{NPZ}$~~  & ~~no error~~ \\
   \hline
   $M_1\in \text{PZ}, ~~M_2\in \text{PZ}$  & ~~$\bar{X}$ on data qubit 1~~ \\
   \hline
   $M_1\in \text{PZ} , M_2\in \text{NPZ}$  & ~~$\bar{X}$ on data qubit 2~~ \\
   \hline
   $M_1\in \text{NPZ} , M_2\in \text{PZ}$  & ~~$\bar{X}$ on data qubit 3~~ \\
\hline
\end{tabular}
\label{tab:GKP 3rep error syndrome}
\end{table}

For the classical 3-qubit repetition code, the qubit with bit-flip error is identified through the measurement outcome of the ancillary qubits, known as the syndrome. The one-to-one correspondence between the syndrome and the single-qubit bit-flip error~\cite{nielsen2001quantum,devitt2013quantum} is summarized in Tab.~\ref{tab:classical 3rep error syndrome}. As an example, if the ancillary qubit $A_1^\prime$ is flipped while $A_2^\prime$ is not, then the states of the data qubits $D_1$ and $D_2$ are different, while the states of the data qubits $D_1$ and $D_3$ are the same. This implies that the data qubit $D_2$ is flipped. 

For the GKP repetition code, the way to identify the bit-flip error through the syndrome is similar. The correspondence between measurement outcomes $\{ M_1, M_2\}$ and the logical Pauli $\bar{X}$ error on different GKP qubits is summarized in Tab.~\ref{tab:GKP 3rep error syndrome}. However, this decoding procedure has a subtle difference from that of the classical repetition code: sometimes a single-qubit Pauli $\bar{X}$ error could be misidentified. As an example, if $u_1'=\sqrt{\pi}\in \text{PZ}$, $u_2'=\sqrt{\pi}/3 \in \text{NPZ}$, $u_3'=\sqrt{\pi}/3 \in \text{NPZ}$, $\alpha_1=\alpha_2=\sqrt{\pi}/3$, then $M_1=2k_1\sqrt{\pi}+u_1'+u_2'+\alpha_1=2k_1\sqrt{\pi}+5\sqrt{\pi}/3 \in \text{NPZ}$, $M_2=2k_2\sqrt{\pi}+u_1'+u_3'+\alpha_2=2k_2\sqrt{\pi}+5\sqrt{\pi}/3 \in \text{NPZ}$, from which we infer that no Pauli $\bar{X}$ error occurs but in fact a Pauli $\bar{X}$ error did occur in the qubit $D_1$. Continuity of the phase space is what makes GKP repetition code different from the classical repetition code~\cite{fukui2017analog}.

Our objective is to calculate the failure probability of 3-qubit GKP repetition code. In contrary to the classical repetition code, we need to reverse the decoding process and impose some conditions to be satisfied instead. For example, if no error occurs, we need $M_1 \in \text{NPZ}$ and $M_2\in \text{NPZ}$ to give the correct identification, and the area outside $M_1 \in \text{NPZ}$ and $M_2\in \text{NPZ}$ must lead to failure. Similar rules apply for other cases. All possible circumstances are summarized as follows:
\begin{itemize}
\item Case 1: If no error occurs $\Rightarrow$we require $M_1\in \text{NPZ}, M_2 \in \text{NPZ}$; 
\item Case 2: If $\bar{X}$ applies on data qubit $D_1$ $\Rightarrow$ we require $M_1\in \text{PZ}, M_2\in \text{PZ}$; 
\item Case 3: If $\bar{X}$ applies on data qubit $D_2$ $\Rightarrow$ we require $M_1\in \text{PZ}, M_2\in \text{NPZ}$; 
\item Case 4: If $\bar{X}$ applies on data qubit $D_3$ $\Rightarrow$ we require $M_1\in \text{NPZ}, M_2\in \text{PZ}$;
\item Case 5: If errors occur on more than one data qubit, with probability $3P_F^2(1-P_F)+P_F^3$  $\Rightarrow$ error correction fails. 
\end{itemize}

Now we calculate the failure probability for the above five cases, the sum of which gives the total failure probability. Consider case 1, there are five constraints needed to be satisfied simultaneously,
\begin{widetext}
\begin{eqnarray}\label{eq:NoErrorDomain}
\text{No Pauli $\bar{X}$ error} &\Rightarrow& \left | u_1'-2m_1\sqrt{\pi}  \right |<\frac{\sqrt{\pi}}{2}, \left | u_2'-2m_2\sqrt{\pi}  \right |<\frac{\sqrt{\pi}}{2}, \left | u_3'-2m_3\sqrt{\pi}  \right |<\frac{\sqrt{\pi}}{2}, \nonumber\\
M_1\in \text{NPZ}, M_2 \in \text{NPZ} &\Rightarrow& \left | u_1'+u_2'+\alpha _1-2n_1\sqrt{\pi}  \right |<\frac{\sqrt{\pi}}{2}, \left | u_1'+u_3'+\alpha _2-2n_2\sqrt{\pi}  \right |<\frac{\sqrt{\pi}}{2},
\end{eqnarray}
where $m_i \in \mathbb{Z}$ and $n_i \in \mathbb{Z}$. 
The probability of success is obtained by integrating the probability distribution of five variables in the domain defined by these five inequalities. However, it is challenging to derive an analytic expression for the success probability, which requires a five-dimensional linear programming. Therefore, a numerical integration method is used instead, which proceeds in two steps. We first fix a point $(u_1',u_2',u_3')$ in the domain defined by the first three inequalities in Eq.~\eqref{eq:NoErrorDomain}, then the success probability at this given point is
\begin{eqnarray}
&&P_\alpha ^1(u_1',u_2',u_3')=\left ( \sum_{n_1}^{}\int_{-\sqrt{\pi}/2+2n_1\sqrt{\pi}-u_1'-u_2'}^{\sqrt{\pi}/2+2n_1\sqrt{\pi}-u_1'-u_2'}f_{q_1'}(\alpha _1) {\rm d} \alpha _1   \right ) \left ( \sum_{n_2}^{}\int_{-\sqrt{\pi}/2+2n_2\sqrt{\pi}-u_1'-u_3'}^{\sqrt{\pi}/2+2n_2\sqrt{\pi}-u_1'-u_3'}f_{q_2'}(\alpha _2) {\rm d} \alpha _2   \right ) \nonumber\\
&&\approx \frac{1}{4} \left [ {\rm erf} \left ( \frac{\frac{\sqrt{\pi}}{2} -u_1'-u_2' }{\tilde{\Delta}}  \right ) - {\rm erf} \left ( \frac{-\frac{\sqrt{\pi}}{2} -u_1'-u_2' }{\tilde{\Delta}} \right )  \right ]  
\left [ {\rm erf} \left ( \frac{\frac{\sqrt{\pi}}{2}-u_1'-u_3' }{\tilde{\Delta}}  \right )- {\rm erf} \left ( \frac{-\frac{\sqrt{\pi}}{2}-u_1'-u_3' }{\tilde{\Delta}}  \right )  \right ],
\end{eqnarray}
where we have only kept one term with $n_1 = n_2 = 0$ in the summation because the contribution from other terms is negligible. Then the failure probability of case 1 is given by integrating the failure probability $1-P_{\alpha }^1(u_1',u_2',u_3')$ over all points satisfying the first three constraints in Eq.~\eqref{eq:NoErrorDomain}, weighted by the probability distribution $F(u_1',u_2',u_3') = F(u_1') F(u_2') F(u_3')$,
\begin{eqnarray}
P_{f,\text{3-rep}}^1 &=& \int_{u_1'\in {\rm NPZ}}^{} \int_{u_2'\in {\rm NPZ}}^{} \int_{u_3'\in {\rm NPZ}}^{} F(u_1',u_2',u_3') \left [ 1-P_{\alpha }^1(u_1',u_2',u_3') \right ] {\rm d} u_1' {\rm d} u_2' {\rm d} u_3'  \nonumber\\ 
&\approx& \int_{u_1'=-\sqrt{\pi}/2 }^{\sqrt{\pi}/2 } \int_{u_2'=-\sqrt{\pi}/2 }^{\sqrt{\pi}/2 } \int_{u_3'=-\sqrt{\pi}/2 }^{\sqrt{\pi}/2 } F(u_1',u_2',u_3')\left [ 1-P_{\alpha }^1(u_1',u_2',u_3') \right ] {\rm d} u_1' {\rm d} u_2' {\rm d} u_3', 
\end{eqnarray}
where we have only kept one term with $m_1 = m_2 = m_3 = 0$ in the summation because the contribution from other terms is negligible.

In a similar way, we can derive the failure probability for case 2 by taking into account the condition that $u_1' \in {\rm PZ}, u_2' \in {\rm NPZ}$ and $ u_3' \in {\rm NPZ}$, 
\begin{eqnarray}
P_{f,\text{3-rep}}^2 &=& \int_{u_1' \in {\rm PZ}}^{} \int_{u_2' \in {\rm NPZ}}^{} \int_{u_3' \in {\rm NPZ}}^{} F(u_1',u_2',u_3')\left [ 1-P_{\alpha }^2(u_1',u_2',u_3') \right ]{\rm d} u_1' {\rm d} u_2' {\rm d} u_3' \nonumber\\ 
&\approx& 2\int_{u_1'=\sqrt{\pi}/2 }^{3\sqrt{\pi}/2 } \int_{u_2'=-\sqrt{\pi}/2 }^{\sqrt{\pi}/2 } \int_{u_3'=-\sqrt{\pi}/2 }^{\sqrt{\pi}/2 } F(u_1',u_2',u_3')\left [ 1-P_{\alpha }^2(u_1',u_2',u_3') \right ] {\rm d} u_1' {\rm d} u_2' {\rm d} u_3',
\end{eqnarray}
where $P_{\alpha }^2(u_1',u_2',u_3')$ is the success probability for a given point $(u_1',u_2',u_3')$ when $M_1\in \text{PZ}$ and $M_2 \in \text{PZ}$,
\begin{eqnarray}
&&P_\alpha ^2(u_1',u_2',u_3')=\left ( \sum_{n_1}^{}\int_{\sqrt{\pi}/2+2n_1\sqrt{\pi}-u_1'-u_2'}^{3\sqrt{\pi}/2+2n_1\sqrt{\pi}-u_1'-u_2'}f_{q_1'}(\alpha _1) {\rm d} \alpha _1   \right ) \left ( \sum_{n_2}^{}\int_{\sqrt{\pi}/2+2n_2\sqrt{\pi}-u_1'-u_3'}^{3\sqrt{\pi}/2+2n_2\sqrt{\pi}-u_1'-u_3'}f_{q_2'}(\alpha _2) {\rm d} \alpha _2   \right ) \nonumber\\
&&\approx \frac{1}{4} \left [ {\rm erf} \left ( \frac{\frac{3\sqrt{\pi}}{2} -u_1'-u_2' }{\tilde{\Delta}}  \right ) - {\rm erf} \left ( \frac{\frac{\sqrt{\pi}}{2} -u_1'-u_2' }{\tilde{\Delta}} \right )  \right ] \left [ {\rm erf}\left ( \frac{\frac{3\sqrt{\pi}}{2}-u_1'-u_3' }{\tilde{\Delta}}  \right ) - {\rm erf} \left ( \frac{\frac{\sqrt{\pi}}{2}-u_1'-u_3' }{\tilde{\Delta}}  \right )  \right ].
\end{eqnarray}
$P_{\alpha }^3(u_1',u_2',u_3')$ and $P_{\alpha }^4(u_1',u_2',u_3')$ can also be calculated by the similar way, and it can be shown that the failure probabilities of cases 3 and 4 are the same as that of the case 2, namely, 
\begin{equation}
P_{f, \text{3-rep}}^2=P_{f, \text{3-rep}}^3=P_{f, \text{3-rep}}^4.
\end{equation}
The failure probability of case 5 is
\begin{equation}
P_{f,\text{3-rep}}^5=3P_F^2(1-P_F)+P_F^3.
\end{equation}
Finally, the total failure probability of the 3-qubit GKP repetition code is
\begin{eqnarray}
P_{f,  \text{3-rep}}(\Delta,\tilde{\Delta}) &=& P_{f,  \text{3-rep}}^1+P_{f,  \text{3-rep}}^2+P_{f,  \text{3-rep}}^3+P_{f,  \text{3-rep}}^4+P_{f,  \text{3-rep}}^5 \nonumber\\ 
&\approx&\int_{u_1'=-\sqrt{\pi}/2 }^{\sqrt{\pi}/2 } \int_{u_2'=-\sqrt{\pi}/2 }^{\sqrt{\pi}/2 } \int_{u_3'=-\sqrt{\pi}/2 }^{\sqrt{\pi}/2 } F(u_1',u_2',u_3')\left [ 1-P_{\alpha }^1(u_1',u_2',u_3') \right ]{\rm d} u_1' {\rm d} u_2' {\rm d} u_3' \nonumber\\ 
&& +6\int_{u_1'=\sqrt{\pi}/2 }^{3\sqrt{\pi}/2 } \int_{u_2'=-\sqrt{\pi}/2 }^{\sqrt{\pi}/2 } \int_{u_3'=-\sqrt{\pi}/2 }^{\sqrt{\pi}/2 } F(u_1',u_2',u_3')\left [ 1-P_{\alpha }^2(u_1',u_2',u_3') \right ]{\rm d}u_1' {\rm d} u_2' {\rm d} u_3' \nonumber\\ 
&&+3P_F^2(1-P_F)+P_F^3.
\end{eqnarray}
\end{widetext}
The last two terms correspond to the failure probability of the classical 3-qubit repetition code, and the first two terms correspond to the failure probability when error occurs on no more than one data qubit, which is what makes GKP repetition code different from the classical repetition code.

\begin{figure}[H]
\includegraphics[width=0.85\columnwidth]{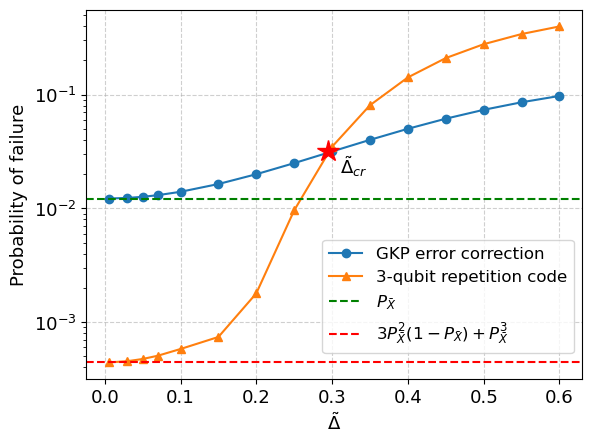}
\caption{Comparison of logical Pauli error rate $P_{f, \text{3-rep}}(\Delta,\tilde{\Delta})$ and $P_F(\Delta,\tilde{\Delta})$ for a fixed $\Delta$, where we choose $\Delta=0.5$ as an example. When $\tilde{\Delta} \to 0$, the failure probability of the 3-qubit GKP repetition code approaches to that of the classical 3-qubit repetition code. There exists a critical value at $\Tilde{\Delta} = \tilde{\Delta}_{\rm cr}$, below which $P_{f, \text{3-rep}}(\Delta,\tilde{\Delta}) < P_F(\Delta,\tilde{\Delta})$ and above which $P_{f, \text{3-rep}}(\Delta,\tilde{\Delta}) > P_F(\Delta,\tilde{\Delta})$. 
}
\label{fig:failureprob 3rep}
\end{figure}

The relation between the failure probability $P_{f, \text{3-rep}}(\Delta,\tilde{\Delta})$ and $\tilde{\Delta}$ for $\Delta=0.5$ is shown in Fig.~\ref{fig:failureprob 3rep}, in which $P_F(\Delta,\tilde{\Delta})$ is also included for comparison. From Fig.~\ref{fig:failureprob 3rep} we can see that
$P_{f, \text{3-rep}}(\Delta,\tilde{\Delta})$ is monotonically decreasing as $\tilde{\Delta}$ decreases. This implies that ancillary qubits with higher quality lead to a lower logical Pauli error rate. When $\tilde{\Delta}\to 0$, the logical Pauli error rate approaches to that of the classical 3-qubit repetition code, namely, 
\begin{eqnarray}
P_{f, \text{3-rep}}(\Delta,\tilde{\Delta}\to 0) &=& 3P_{\bar{X}}^2(\Delta)[1-P_{\bar{X}}(\Delta)]+P_{\bar{X}}^3(\Delta) \nonumber\\
&=& P_{f, \text{3-rep}}^{\rm class}(\Delta).
\end{eqnarray}
This can be understood as follows. The probability distribution of $u_1',u_2',u_3',\alpha_1$ and $\alpha_2$ are determined by $\tilde{\Delta}$,
\noindent and when $\tilde{\Delta}\to 0$ the distribution is highly localized and is close to a $\delta$ function. As a result, the probability of misidentifying no error and single-qubit Pauli $\bar{X}$ errors is almost zero. When $\tilde{\Delta}$ is large, the failure probability $P_{f, \text{3-rep}}(\Delta,\tilde{\Delta})$ is greater than $P_F(\Delta,\tilde{\Delta})$. However, the former decreases faster than the latter as $\Tilde{\Delta}$ decreases. There exists a critical value for $\Tilde{\Delta}$, denoted as $\tilde{\Delta}_{\rm cr}$, such that $P_{f, \text{3-rep}}(0.5,\tilde{\Delta}_{\rm cr}) = P_F(0.5,\tilde{\Delta}_{\rm cr})$, and from Fig.~\ref{fig:failureprob 3rep} we can see that $\tilde{\Delta}_{\rm cr} \approx 0.3$. When $\Tilde{\Delta} < \tilde{\Delta}_{\rm cr}$, the logical Pauli error rate of the 3-qubit GKP repetition code is lower than that of the GKP code with noisy ancillary qubits. The concatenation with repetition code therefore shows its advantage in this regime. Furthermore, the failure probability $P_{f, \text{3-rep}}(\Delta,\tilde{\Delta})$ can be even lower than that of the GKP code with ideal ancillary qubits, namely, $P_{f, \text{3-rep}}(0.5,\tilde{\Delta}) < P_{\bar{X}}(0.5)$. 

\subsection{Concatenation with $n$-qubit repetition code}\label{sec:4B}

\begin{figure}
\includegraphics[width=1.0\columnwidth]{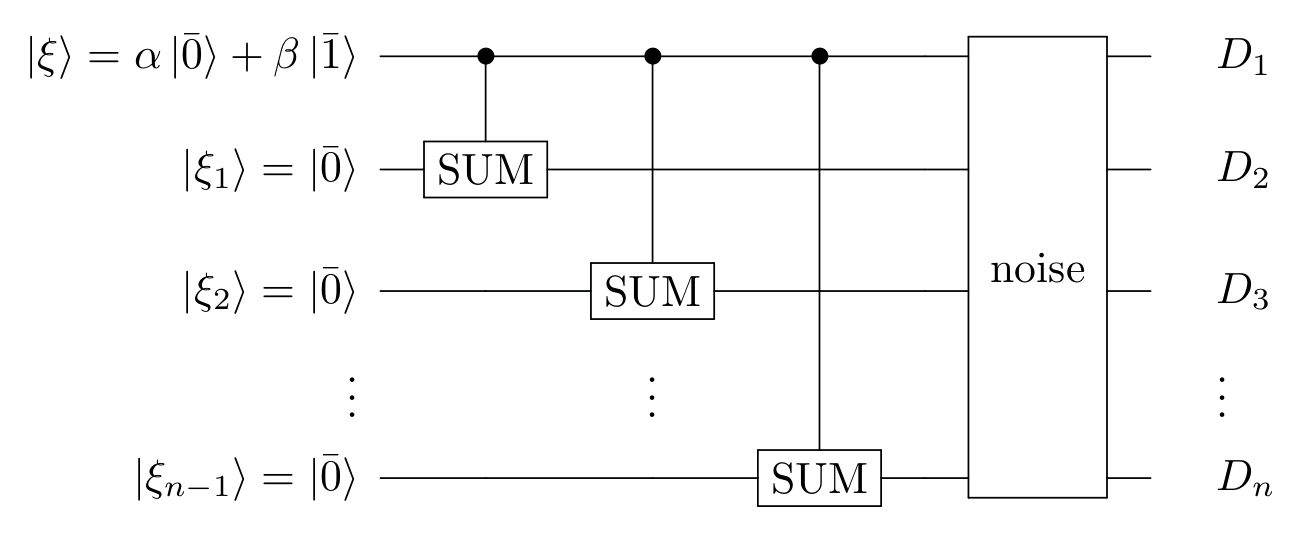}
\caption{Encoding circuit for $n$-qubit GKP repetition code. The $n$ GKP states before encoding are assumed to be ideal. After encoding, $n$ data qubits entangle with each other. Then a physical GKP $n$-qubit repetition code state is constructed by coherently superposing the ideal GKP states undergoing random displacements.}
\label{fig:encode nrep}
\end{figure}

In the theory of quantum error correction, introducing more qubits would allow more errors to be corrected and therefore achieve a lower logical error rate~\cite{PhysRevA.52.R2493,gottesman1997stabilizer,terhal2015quantum}. In this section, we generalize the previous scheme and concatenate the GKP code with $n$-qubit repetition code, with $n$ an odd integer. We are gonna to show that increasing the size of the GKP repetition code can further reduce the logical Pauli error rate, though one needs to prepare ancillary GKP qubits with higher quality.

For the classical $n$-qubit repetition code~\cite{nielsen2001quantum,devitt2013quantum}, $n$ physical qubits are introduced to encode one logical qubit, in particular, the single-qubit state $\alpha \ket{0}+\beta \ket{1}$ is encoded as follows:
\begin{equation}
\ket{\psi}=\alpha\ket{0}+\beta\ket{1}\to \ket{\bar{\psi}} = \alpha\ket{00 \cdots 0}+\beta\ket{11 \cdots 1}.
\end{equation}
The $n$-qubit repetition code is able to correct any $m$-qubit bit-flip error, with $1 \le m \le (n-1)/2$.  Denote the bit-flip error rate of a single physical qubit as $p$, then the failure
\begin{table}[H]
\caption{Correspondence between syndromes and single-qubit bit-flip errors for the $n$-qubit GKP repetition code.}
\centering
\begin{tabular}{c|c}
\hline \hline
   {\bf Measurement outcome} & {\bf Error} \\
    \hline
   $M_1, M_2, \cdots, M_{n-1}\in \text{NPZ}$  & no error \\
   \hline
   $M_1, M_2, \cdots, M_{n-1}\in \text{PZ}$  & data qubit 1 \\
   \hline
   $M_1\in PZ, M_2, \cdots, M_{n-1}\in \text{NPZ}$  & data qubit 2 \\
   \hline
   $M_1\in NPZ, M_2\in PZ, M_3, \cdots, M_{n-1}\in \text{NPZ}$  & data qubit 3 \\
   \hline
   $\cdots \cdots$  &  $\cdots \cdots$ \\
   \hline
   $M_1\in NPZ, M_2, \cdots, M_{n-1}\in \text{PZ}$  & data qubit 1,2 \\
   \hline
   $M_1\in PZ, M_2\in NPZ, M_3, \cdots, M_{n-1}\in \text{PZ}$  & data qubit 1,3 \\
   \hline
   $\cdots \cdots$   & $\cdots \cdots$ \\
   \hline
   $M_1,M_2\in NPZ, M_3, \cdots, M_{n-1}\in \text{PZ}$  & ~data qubit 1,2,3 \\
   \hline
   $\cdots \cdots$ & $\cdots \cdots$ \\
   \hline
\end{tabular}
\label{tab:GKP nrep code}
\end{table}
\noindent probability of the $n$-qubit repetition code is given by
\begin{equation}
P_{f, \text{n-rep}}^{\rm class}=\sum_{i=(n+1)/2 }^{n}C_{n}^i \, p^i(1-p)^{n-i}.
\end{equation}
There are $C_n^0+C_n^1+...+C_n^{(n-1)/2}=2^{n-1}$ possibilities that up to $(n-1)/2$ quibts are flipped, i.e., $2^{n-1}$ correctable errors. Therefore, $2^{n-1}$ syndromes are needed to decode these errors, which implies $(n-1)$ ancillary GKP qubits are required to perform syndrome measurement.

\begin{figure*}
\includegraphics[width=1.55\columnwidth]{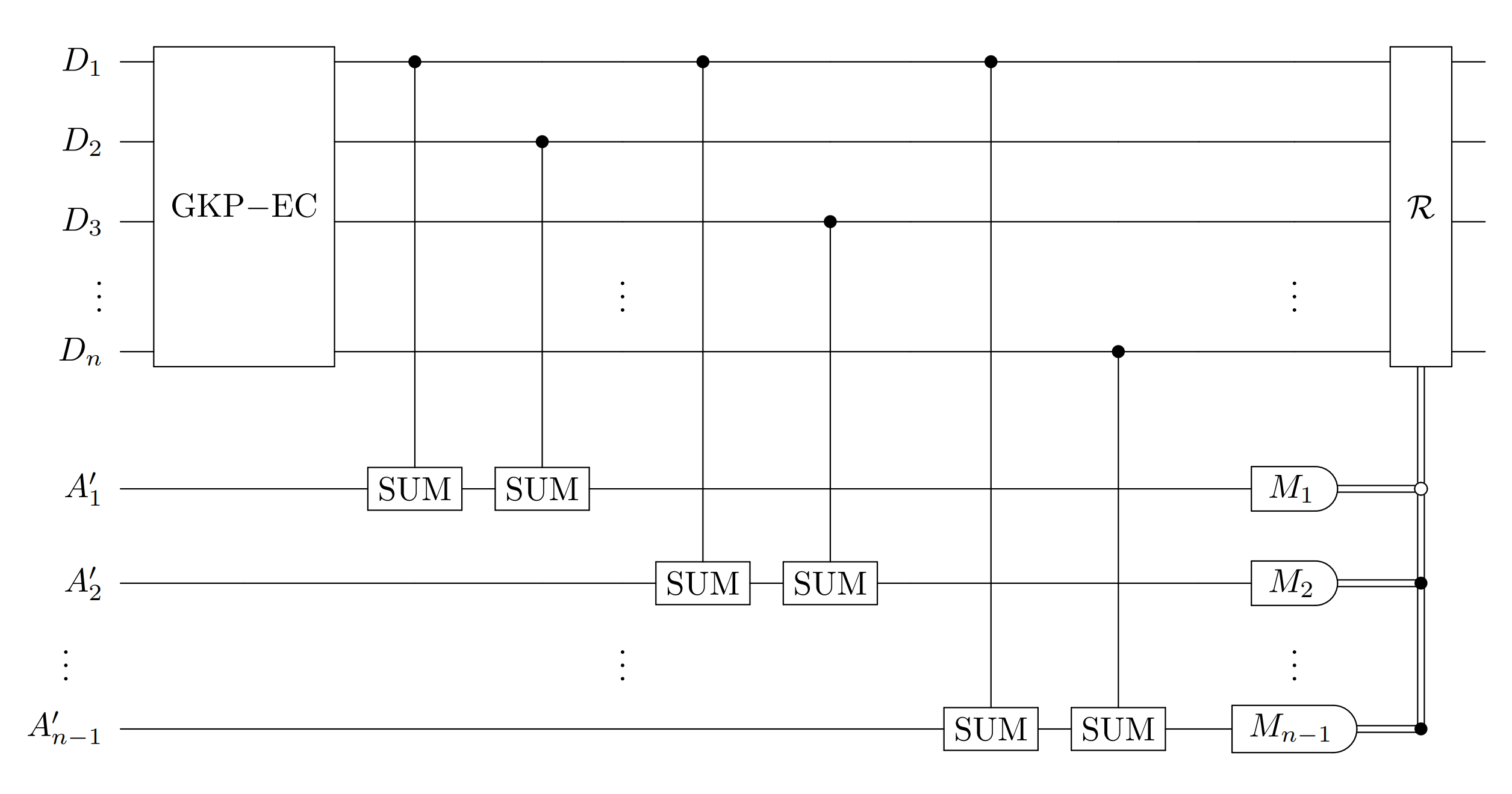}
\caption{
Quantum error correction circuit for $n$-qubit GKP repetition code. It consists of one round of GKP error correction, syndrome measurement of repetition code, and recovery operation according to the measurement outcomes $\{M_i\}_{i=1}^{n-1}$. Here $\{ D_i\}_{i=1}^n$ denote the data qubits and $\{ A_i^\prime\}_{i=1}^{n-1}$ denote the ancillary qubits introduced to perform syndrome measurement. }
\label{fig:nrep code}
\end{figure*}

We now concatenate the GKP code with the $n$-qubit repetition code. The quantum circuit of encoding is shown in Fig.~\ref{fig:encode nrep}, where we assume all input GKP states are ideal. Before encoding, the first GKP qubit is prepared in the state $\alpha \ket{\bar{0}} + \beta \ket{\bar{1}}$, and all other GKP qubits are prepared in the state $\ket{\bar{0}}$. After the encoding procedure, namely, the application of ($n-1$) SUM gates, the $n$ GKP qubits become entangled with each other,
\begin{equation}
\ket{\xi,\xi_1,\xi_2,\cdots,\xi_{n-1}} \to \ket{\bar{\psi}_n} = \alpha\ket{\bar{0}\bar{0}\bar{0} \cdots \bar{0}}+\beta\ket{\bar{1}\bar{1}\bar{1} \cdots \bar{1}}.
\end{equation}
We then construct a physical GKP repetition code state by coherently superposing the randomly displaced ideal GKP repetition code states, namely
\begin{eqnarray}
\ket{\tilde{\Psi}_n} &=& \int {\rm d} u_1 {\rm d} v_1\cdots {\rm d} u_n {\rm d} v_n \, \eta(u_1,v_1)\cdots \eta(u_n,v_n) \nonumber\\
&& \times\exp \bigg\{ -i \sum_{k=1}^n (u_k \hat{p}_k - v_k \hat{q}_k) \bigg\} \ket{\bar{\psi}_n}.
\end{eqnarray}
We assume that the displacement in each ideal GKP qubit is independent and follows the same probability distribution.

The quantum circuit of error correction is shown in Fig.~\ref{fig:nrep code}, which is a direct generalization to that of the 3-qubit GKP repetition code, and the procedure of quantum error correction is also similar. The residual displacements of the $n$ data qubits after the GKP error correction are denoted as $\{ u_i^\prime \}_{i=1}^n$, and their probability distribution is given by Eq.~\eqref{eq:udot distribution}. The displacement errors of the $(n-1)$ ancillary qubits for syndrome measurement are denoted as $\{ \alpha_i \}_{i=1}^{n-1}$, and their probability distribution is given by Eq.~\eqref{eq:ancilla distribution}. The outcomes of the syndrome measurement are given by
\begin{eqnarray}
M_i = 2 k_i \sqrt{\pi} + u_1^\prime + u_{i+1}^\prime + \alpha_i, ~~ i = 1, 2, \cdots, n-1,
\end{eqnarray}
with $k_i \in \mathbb{Z}$. Similarly, there is a one-to-one correspondence between the syndromes and correctable errors, which is summarized in Tab.~\ref{tab:GKP nrep code}. 

Using the same method, we can calculate the failure probability of the $n$-qubit GKP repetition code (see Appendix ~\ref{sec:appC} for details),
\begin{widetext}
\begin{eqnarray}\label{eq:failureprob nrep}
&&P_{f, \text{n-rep}}=\int_{u_1'=-\sqrt{\pi}/2}^{\sqrt{\pi}/2} \cdots \int_{u_n'=-\sqrt{\pi}/2}^{\sqrt{\pi}/2} \bigg[ \prod_{i=1}^n F(u_i^\prime) \bigg] \big[1-P_{\alpha }^1(u_1', \cdots, u_n') \big] {\rm d} u_1' \cdots {\rm d} u_n' \nonumber\\
&&+\sum_{m=1}^{\frac{n-1}{2} }C_n^m 2^m \int_{u_1'=\sqrt{\pi}/2}^{3\sqrt{\pi}/2} \cdots \int_{u_m'=\sqrt{\pi}/2}^{3\sqrt{\pi}/2}\int_{u_{m+1}'=-\sqrt{\pi}/2}^{\sqrt{\pi}/2} \cdots \int_{u_n'=-\sqrt{\pi}/2}^{\sqrt{\pi}/2}\bigg[ \prod_{i=1}^n F(u_i^\prime) \bigg] \big[1-P_{\alpha }^{m+1}(u_1', \cdots, u_n') \big] {\rm d} u_1' \cdots {\rm d} u_n' \nonumber\\
&&+\sum_{i=\frac{n+1}{2} }^{n} C_n^i P_F^i(1-P_F)^{n-i}.
\end{eqnarray}
The last summation represents the failure probability of the classical $n$-qubit repetition code, and the first two terms represent the failure probability when no more than $(n-1)/2$ data qubits are flipped. Here $P_{\alpha}^{s}(u_1',...,u_n')$ $(1\le s \le \frac{n+1}{2})$ is the success probability for a fixed point $(u_1',...,u_n')$ when $(s-1)$ qubits are flipped. Although there are many possible ways that $(s-1)$ qubits are flipped, it can be shown that the success probabilities corresponding to these cases are the same. The expressions for $P_{\alpha}^{s}(u_1',...,u_n')$ 
are given by
\begin{equation}
\begin{split}
P_{\alpha}^1(u_1', \cdots, u_n')=&\frac{1}{2^{n-1}}\prod_{k=2}^{n}\left [ {\rm erf} \left ( \frac{\sqrt{\pi}/2-u_1'-u_k' }{\tilde {\Delta}}  \right )-{\rm erf} \left ( \frac{-\sqrt{\pi}/2-u_1'-u_k' }{\tilde {\Delta}} \right )   \right ], \\
P_{\alpha}^2(u_1', \cdots, u_n')=&\frac{1}{2^{n-1}}\prod_{k=2}^{n}\left [ {\rm erf} \left ( \frac{3\sqrt{\pi}/2-u_1'-u_k' }{\tilde {\Delta}}  \right )-{\rm erf} \left ( \frac{\sqrt{\pi}/2-u_1'-u_k' }{\tilde {\Delta}} \right )   \right ], \\
P_{\alpha}^s(u_1', \cdots, u_n')=&\frac{1}{2^{n-1}}\prod_{k_1=2}^{s-1}\left [ {\rm erf} \left ( \frac{5\sqrt{\pi}/2-u_1'-u_{k_1}' }{\tilde {\Delta}}  \right )-{\rm erf} \left ( \frac{3\sqrt{\pi}/2-u_1'-u_{k_1}'}{\tilde {\Delta}} \right )   \right ]\\
& \times \prod_{k_2=s}^{n}\left [ {\rm erf} \left ( \frac{3\sqrt{\pi}/2-u_1'-u_{k_2}' }{\tilde {\Delta}}  \right )-{\rm erf} \left ( \frac{\sqrt{\pi}/2-u_1'-u_{k_2}' }{\tilde {\Delta}} \right )   \right ], ~\text{for} ~ 3\le s\le \frac{n+1}{2}.
\end{split}
\end{equation}
\end{widetext}

\begin{figure}
\includegraphics[width=0.9\columnwidth]{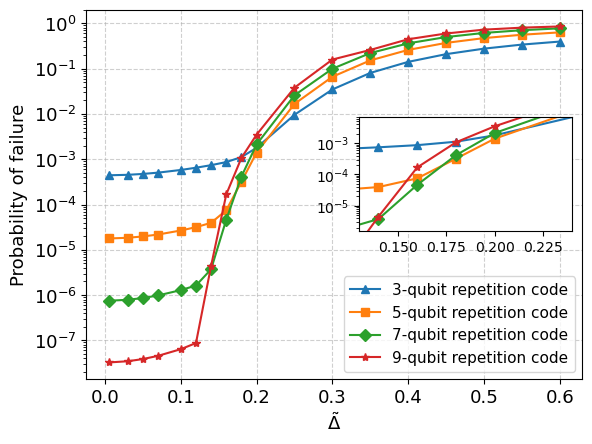}
\caption{ Comparison of logical Pauli error rate $P_{f,\text{n-rep}}(\Delta,\tilde{\Delta})$  of $n$-qubit GKP repetition codes for $n$ from 3 to 9, with $\Delta=0.5$ as an example. The inset shows the location of various critical values. }
\label{fig:failureprob nrep}
\end{figure}

We calculate the failure probability $P_{f, \text{n-rep}}(\Delta,\tilde{\Delta})$ for the GKP repetition code with the number of data quabits up to $n = 9$. The results for a fixed $\Delta$ are shown in Fig.~\ref{fig:failureprob nrep}, in which we choose $\Delta=0.5$ as an example. Firstly, it can be seen that the failure probability $P_{f, \text{n-rep}}(\Delta,\tilde{\Delta})$ for all $n$ monotonically decreases as $\tilde{\Delta}$ decreases. This implies that ancillary qubits with higher quality lead to a lower logical Pauli error rate. In the limit of $\tilde{\Delta}\to 0$, the logical Pauli error rate approaches to that of the classical $n$-qubit repetition code, namely,
\begin{eqnarray}
P_{f, {\text{n-rep}}}(\Delta,\tilde{\Delta} \to 0) 
&=& \sum_{i=\frac{n+1}{2} }^{n} C_n^i P_{\bar{X}}^i(\Delta) [1-P_{\bar{X}}(\Delta)]^{n-i} \nonumber\\
&=& P_{f, \text{n-rep}}^{\rm class}(\Delta).
\end{eqnarray}
The second observation is that when $\tilde{\Delta}$ is sufficiently large, the logical Pauli error rate increases as the size of the code increases; when $\tilde{\Delta}$ is sufficiently small, the logical Pauli error rate decreases as the size of the code increases. This implies that the concatenation of GKP code with repetition code can reduce the logical Pauli error rate under the condition that the quality of the ancillary qubit is sufficiently high. Figure~\ref{fig:failureprob nrep} indicates that there exists some threshold for $\tilde{\Delta}$, below which the concatenation shows advantages. However, the location of the threshold is not sharp. Define the critical noise variance $\tilde{\Delta}_{nm}^2$ as the variance of the ancillary qubit when the $n$-qubit GKP repetition code and the $m$-qubit GKP repetition code have the same logical Pauli error rate for a fixed $\Delta$. From Fig.~\ref{fig:failureprob nrep} it can be seen that $\tilde{\Delta}_{97} < \tilde{\Delta}_{75} < \tilde{\Delta}_{53}$. They are close but not the same. When $\tilde{\Delta} > \tilde{\Delta}_{53}$, we have $P_{f, \text{9-rep}}>P_{f, \text{7-rep} }>P_{f, \text{5-rep}}>P_{f, \text{3-rep}}$. Therefore, $\tilde{\Delta}_{53}^2$ can be considered as the minimal noise variance that the concatenation with repetition code is completely useless. When $\tilde{\Delta} < \tilde{\Delta}_{97}$, we have $P_{f, \text{9-rep}} < P_{f, \text{7-rep} } < P_{f, \text{5-rep}} < P_{f, \text{3-rep}}$. Therefore, $\tilde{\Delta}_{97}^2$ is the noise variance that one needs to achieve in order to realize the power of repetition code concatenation with at least 9 GKP qubits.

\begin{figure}
\includegraphics[width=0.9 \columnwidth]{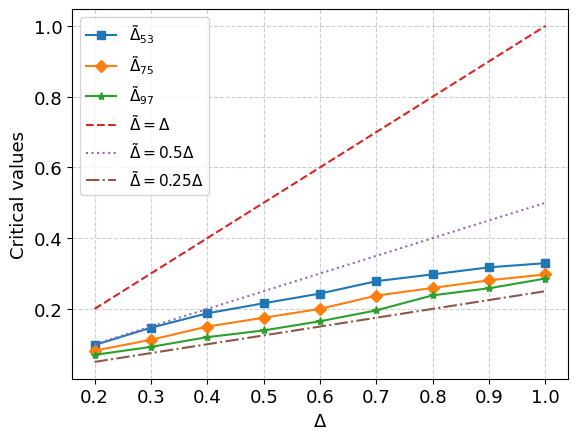}
\caption{Relation between  $\tilde{\Delta}_{nm}$ and $\Delta$. The ratio $\tilde{\Delta}_{nm}/\Delta$ is upper bounded by 0.5 and lower bounded by 0.25 for the code size $n$ from 3 to 9.}
\label{fig:thresholds}
\end{figure}

The critical noise variance $\tilde{\Delta}_{nm}^2$ depends on the noise variance of the data qubits. The relation between $\tilde{\Delta}_{nm}$ and $\Delta$ is shown in Fig.~\ref{fig:thresholds}. We can see that $\tilde{\Delta}_{nm}$ increases monotonically as $\Delta$ increases. This implies that a lower-quality GKP repetition code requires lower quality ancillary qubits to achieve its advantage. This is rather surprising and counter intuitive. However, one should keep in mind that this does not imply that a low-quality GKP repetition code is preferred in the experimental realization. This is because one also needs to take into account the displacement error in momentum space, which we will discuss later in Sec.~\ref{sec:5}. 
From Fig.~\ref{fig:thresholds} it can be seen that the relation between $\tilde{\Delta}_{nm}$ and $\tilde{\Delta}$ is almost linear, we therefore define an approximate ratio $\tilde{\Delta}_{nm}/\Delta$ (or an average ratio).  The ratio depends on the size of the code, and is upper bounded by 0.5 and lower bounded by 0.25 for the code size that we consider. This means if we choose $\tilde{\Delta} = \Delta$, the logical Pauli error rate increases as the size of the code increases, implying that concatena-
\noindent tion with repetition code is useless; while if we choose $\tilde{\Delta} = 0.25 \Delta$, the logical Pauli error rate decreases as the size of the code increases, implying that concatenation with repetition code with at least 9 GKP qubits is useful.

We are not able to calculate the failure probability for arbitrarily large $n$ since it involves a very high-dimensional integral, which is a rather challenging task. Based on the results with $n$ up to nine, we conjecture that there exists a nonzero threshold for $\tilde{\Delta}$ such that for sufficiently small $\Delta$ the logical Pauli error rate can be exponentially suppressed by increasing the size of the code. This threshold can be calculated by using the Monte Carlo simulation and we leave it for future work. 

\subsection{Comparison with no GKP error correction}\label{sec:4C}

Although the GKP error correction increases the probability of Pauli $\bar{X}$ error for all values of $\tilde{\Delta}$, it narrows down the error distribution of the GKP state when $\tilde{\Delta}< \Delta$ such that the concatenation with repetition code is advantageous. However, the GKP error correction requires the same number of ancillary GKP qubits as the data qubits. A question arises as to whether the GKP error correction is necessary in order to reduce the logical Pauli error rate. If the GKP error correction is not necessary, then we only need to supply ancillary GKP qubits for syndrome measurement and therefore can save a substantial amount of physical resources. 

\begin{figure}[H]
\includegraphics[width=0.9\columnwidth]{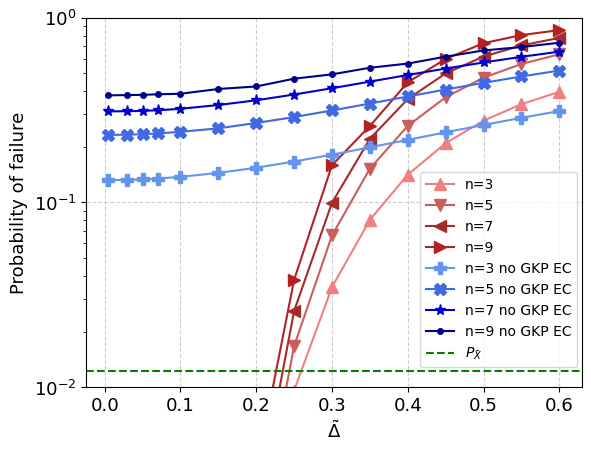}
\caption{ Comparison of failure probabilities for GKP repetition codes without (blue) and with (red) one round of GKP error correction before concatenation. Here we choose $\Delta=0.5$. }
\label{fig:failureprob no SUM}
\end{figure}

To calculate the failure probability of GKP repetition code without one round of GKP error correction, we only need to replace 
$\{ u_i^\prime \}$ with distribution $F(u_1',...,u_n')$ in Eq.~\eqref{eq:failureprob nrep} by $\{ u_i \}$ with distribution $f_q(u_1,...,u_n)$ given by Eq.~\eqref{eq:intrinsic error distribution}, and replace $P_F$ by $P_{\bar{X}}$, where $\{ u_i \}$ are displacement errors of $n$ data qubits. The result is then given by
\begin{widetext}
\begin{eqnarray}
&&P_{f, \text{n-rep}}^{\, \prime} (\Delta,\tilde{\Delta}) = \int_{u_1=-\sqrt{\pi}/2}^{\sqrt{\pi}/2} \cdots \int_{u_n=-\sqrt{\pi}/2}^{\sqrt{\pi}/2}f_{q}(u_1, \cdots, u_n)[1-P_{\alpha }^1(u_1, \cdots, u_n)] {\rm d} u_1{\rm d} u_n \nonumber\\
&&+\sum_{m=1}^{\frac{n-1}{2} }C_n^m 2^m \int_{u_1=\sqrt{\pi}/2}^{3\sqrt{\pi}/2} \cdots \int_{u_m=\sqrt{\pi}/2}^{3\sqrt{\pi}/2}\int_{u_{m+1}=-\sqrt{\pi}/2}^{\sqrt{\pi}/2} \cdots \int_{u_n=-\sqrt{\pi}/2}^{\sqrt{\pi}/2}f_{q}(u_1, \cdots, u_n)[1-P_{\alpha }^{m+1}(u_1, \cdots, u_n)] {\rm d} u_1 \cdots {\rm d} u_n \nonumber\\
&&+\sum_{i=\frac{n+1}{2} }^{n} C_n^iP_{\bar{X}}^i(1-P_{\bar{X}})^{n-i}.
\end{eqnarray}
\end{widetext}
We calculate the failure probability $P_{f, \text{n-rep}}^{\, \prime}(\Delta,\tilde{\Delta})$ for the GKP repetition code with the number of data qubits up to $n = 9$. The results for a fixed $\Delta$ are shown in Fig.~\ref{fig:failureprob no SUM}, in which we choose $\Delta=0.5$ as an example. We also plot $P_{f, \text{n-rep}}(\Delta,\tilde{\Delta})$ in Fig.~\ref{fig:failureprob no SUM} for comparison. We can see that GKP repetition code without one round of GKP error correction cannot reduce the logical Pauli error rate even when $\tilde{\Delta} \rightarrow 0$, and increasing the size of the code leads to a higher logical Pauli error rate. 
We have confirmed that this is true for $\Delta\ge 0.2$, and we expect that this should also be the case when $\Delta<0.2$. Therefore, one round of GKP error correction before concatenation is necessary. This can be understood in an intuitive way as follows. In the repetition code without GKP error correction, the distribution of $u_1,u_2,...,u_n$ depends only on $\Delta$. If we want to reduce the logical Pauli error rate using repetition code, we need to replace the distribution of $u_1,u_2,...u_n$ with a narrower distribution. Repetition code with GKP error correction achieves this by replacing the distribution of $u_1,u_2,...,u_n$ with a narrower distribution of $u_1',u_2',...,u_n'$ when $\tilde{\Delta}\ll \Delta$.

\section{Biased noise corrected by GKP repetition code}\label{sec:5}

Until now we have only considered correcting displacement errors in position space and neglected those in momentum space. Therefore, previous results are only valid in the limit of no errors in momentum space. However, a physical GKP state does have noise in both position and
\noindent momentum spaces. In this section, we take into account the momentum displacement error and assume a biased noise model, namely, with unequal position and momentum noise, and introduce a GKP repetition code to suppress the logical error.

\begin{figure}
\includegraphics[width=1.05\columnwidth]{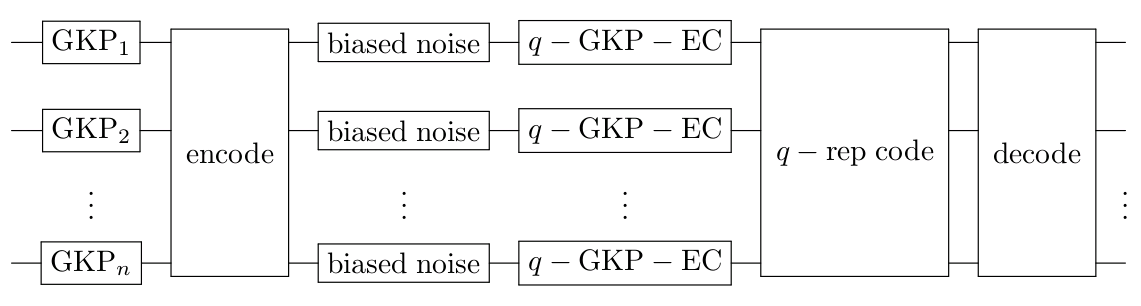}
\caption{ Scheme to correct biased noise using GKP repetition code. Ideal GKP repetition code is first generated by injecting ideal GKP states into the encoding circuit. Biased noise is then imposed to the ideal GKP repetition code to produce a physical GKP repetition code. The error correction consists of one round of GKP error correction, syndrome measurement on repetition code and decoding. }
\label{fig:whole design}
\end{figure}

The scheme of correcting biased noise using GKP repetition code is schematically shown in Fig.~\ref{fig:whole design}, where ``q-GKP-EC" represents GKP error correction in position space, ``q-rep code" represents error correction in position space by concatenating with repetition code. The GKP states before encoding are assumed to be ideal. Biased noise is imposed to the data qubits after encoding, with the error probability distribution given by Eq.~\eqref{eq:biased intrinsic error distribution}.
By choosing $r>1$, the error in momentum space is suppressed at the expense of amplifying the error in position space. Fortunately, this is not a trouble because the displacement error in position space can be efficiently corrected by concatenating the GKP code with repetition code.  

However, it should be noted that further error correction in position space will contaminate the momentum quadrature. From the transformation rule of the SUM gate given by Eq.~\eqref{eq:SUM rule}, the momentum displacement error of the ancillary qubits can propagate to the momentum space of the data qubits, therefore the variance of the error distribution in momentum space will be amplified. The initial noise variance of the physical GKP state in momentum space is $(\Delta/r)^2$. After one round of GKP error correction, the momentum displacement error of the ancillary GKP qubit propagates to the data qubit, resulting in a noise variance $(\Delta/r)^2+\tilde{\Delta}^2$. Concatenation with repetition code will further increase the noise in momentum space because of the sequential application of SUM gates during the syndrome measurement. By concatenating with an $n$-qubit repetition code, the noise variance of the first data qubit in momentum space becomes $(\Delta/r)^2+n\tilde{\Delta}^2$ since it couples with $(n-1)$ ancillary qubits via the SUM gate; while the noise variance of all other data qubits in momentum space becomes $(\Delta/r)^2+2\tilde{\Delta}^2$ since each of them couples with only one ancillary qubit.  

\begin{figure}
\includegraphics[width=0.9\columnwidth]{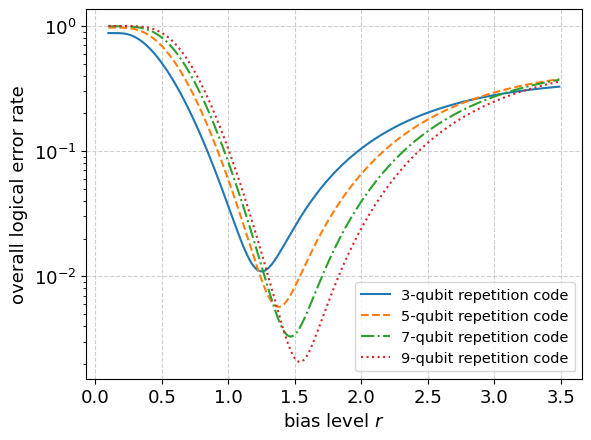}
\caption{Relation between $P_{\rm fail}$ and $r$, with $\Delta=0.5,~\tilde{\Delta}=0$ and $n=3,5,7,9$. For a fixed code size $n$, $P_{\text{fail}}$ first decreases to a minimum and then increases, and the bias level corresponding to the minimal failure probability is the optimal bias level $r_{\text{opt}}$. The optimal bias level $r_{\rm opt}$ increases as the size of the code increases. }
\label{fig:biased noise}
\end{figure}

The logical information is protected when the momentum displacement is in NPZ and the correction of the position displacement using the GKP repetition code succeeds. The overall logical error rate after the error correction is given by
\begin{equation}\label{eq:ERtotal}
\begin{split}
&P_{\rm fail}=1- \bigg[ 1-P_{\bar{Z}}\left ( \sqrt{(\Delta/r)^2+2\tilde{\Delta}^2}  \right )  \bigg]^{n-1} \\
& \times \bigg[ 1-P_{\bar{Z}}\left ( \sqrt{(\Delta/r)^2+n\tilde{\Delta}^2}  \right )  \bigg] \bigg[ 1-P_{f, \text{n-rep}}(r\Delta,\tilde{\Delta}) \bigg] ,
\end{split}
\end{equation}
where the expression for $P_{\bar{Z}}(\Delta)$ is the same as $P_{\bar{X}}(\Delta)$, which is given by Eq.~\eqref{logical error rate X}. 

The contribution from momentum displacement in Eq.~\eqref{eq:ERtotal} scales approximately like $(1 - P_{\bar{Z}})^n$, which exponentially decreases with increasing $n$ since $P_{\bar{Z}}$ is positive and smaller than one. In the case where $P_{f, \text{n-rep}}(r\Delta,\tilde{\Delta})$ increases with increasing $n$, the logical error rate $P_{\rm fail}$ becomes higher for a larger code, indicating that concatenation with repetition code shows no advantages. Therefore, in order to exploit the power of code concatenation, $\tilde{\Delta}$ has to be at least smaller than the upper bound of those critical values, namely, $\tilde{\Delta}_{53}$. This imposes a minimal requirement for the variance of the ancillary GKP quibts. However, the actual condition needed to be satisfied is generally more stringent. The failure probability $P_{f, \text{n-rep}}(r\Delta,\tilde{\Delta})$ has to decrease fast enough with increasing $n$, so that the decreasing of the contribution from the momentum displacement can be compensated and the logical error rate $P_{\rm fail}$ decreases with increasing $n$. The calculation of the exact threshold for $\tilde{\Delta}$ is challenging and we leave it for future work. We conjecture that there exists a nonzero threshold for $\tilde{\Delta}$. 

As an indirect evidence for our conjecture, we show that for ideal ancillary GKP qubits, i.e., $\tilde{\Delta} = 0$, the code concatenation can show advantages in reducing the logical error rate. Note that the error rate $P_{\rm fail}$ is also a function of the bias level $r$. When $r$ is small, the displacement error from momentum space dominates; while when $r$ is large, the displacement error from position space dominates. It is expected that the error correction achieves its best performance in the intermediate regime where we can find an optimal bias level $r_{\rm opt}$. The relation between $P_{\rm fail}$ and $r$ is plotted in Fig.~\ref{fig:biased noise} for $\Delta = 0.5$ and $\tilde{\Delta} = 0$. It is evident that, for every $n$, the logical error rate has a minimum corresponding to the optimal bias level $r_{\rm opt}$. In addition, the optimal bias level increases when the code size increases. Most importantly, the minimal logical error rate is lower for a larger repetition code, showing advantages of concatenating the GKP code with repetition code. However, one should note that $\Delta$ has a threshold when $\tilde{\Delta} = 0$, above which the concatenation with repetition code shows no advantages. The threshold is estimated to be $0.599\times \sqrt{2}\approx 0.847$ in Ref.~\cite{stafford2022biased} (Note that the variance of error distribution of GKP state in our work is twice of that in Ref.~\cite{stafford2022biased}).

\section{Conclusions and outlooks}\label{sec:6}

We study the concatenation of GKP code with repetition code to correct biased random displacement errors with noisy ancillary GKP qubits. The error correction procedure consists of one round of GKP error correction, concatenation with repetition code, syndrome measurement and recovery operation. The purpose of the GKP error correction is to correct displacement errors before concatenation to alleviate the heavy burden of the repetition code. 

We find that, after the GKP error correction, the probability of the logical Pauli $\bar{X}$ error with noisy ancillary qubits always increases as compared to the case where ideal ancillary qubits are used. This is expected because the displacement errors from the noisy ancillary qubits can propagate to the data qubits. We then concatenate the GKP code with repetition code to suppress the logical Pauli $\bar{X}$ error. We find that there exists a critical value for the noise variance of the ancillary qubits, below which the logical Pauli $\bar{X}$ error rate decreases as the size of the repetition code increases; and there is a slightly different critical value for the noise variance of the ancillary qubits, above which the logical Pauli $\bar{X}$ error rate increases as the size of the repetition code increases. These critical values for the noise variance of the ancillary qubits are dependent on the noise variance of the data qubits, and they increases monotonically as the noise variance of the data qubits increases. Their ratio is lower bounded by $1/16$ and upper bounded by 1. This means if we take $\tilde{\Delta}^2 = 1/16 \Delta^2$, the logical Pauli $\bar{X}$ error can be efficiently suppressed by concatenating with repetition code; while if we take $\tilde{\Delta}^2 = \Delta^2$, the logical Pauli $\bar{X}$ error rate increases as the code size increases, at least for repetition codes with the number of data qubits up to nine. We also show that the GKP error correction before concatenation with repetition code is necessary, otherwise the logical Pauli $\bar{X}$ error rate cannot be reduced even for ideal ancillary GKP qubits

We then use the GKP repetition code to correct biased noise, for which the random displacement errors in momentum space are assumed to be smaller than that in position space, and therefore no further concatenation is introduced to correct them. We conjecture that if the noise variance of the ancillary qubit is below a threshold and $\Delta$ is also below a threshold estimated to be 0.847, the concatenation with repetition code can efficiently reduce the overall logical error rate. A GKP repetition code with more data qubits leads to a lower overall logical error rate, albeit with a higher level of noise bias.

Although we have taken into account the effects of noisy ancillary GKP qubits, there are still some assumptions needed to be relaxed in future work. For example, quantum operations (SUM gates and recovery operations) are assumed to be ideal, measurement is assumed to be unbiased, and most importantly, the generation of the encoded states of GKP repetition code is not well studied. Actually, the $n$ input GKP states before encoding are non-ideal states, so the degree of squeezing of the $n$ data qubits after encoding are not the same. However, we use an encoded state with equal degree of squeezing only to facilitate calculations, and focus on the relation between performance of the error correction and the quality of the ancillary qubits. Hence, our work only provides a lower limit of the logical error rate of the error correction code. 
There are some other aspects needed to further explore. It is interesting and important to check whether there is a non-infinite squeezing threshold for the ancillary GKP qubits for an arbitrarily large GKP repetition code. This can be estimated for example by using the Monte Carlo simulation. In addition, one needs to consider the physical encoded states that can be efficiently prepared in the experiment, and take into account the imperfect Homodyne measurement and SUM gate. 
~\\\\ 
{\bf Acknowledgements:} This work is supported by the Fundamental Research Funds for the Central Universities, HUST (Grant No. 5003012068).

\newpage
\appendix
\vspace{0.5cm}

\begin{widetext}

\section{Physical GKP state}\label{sec:appA}
By discussing some examples, we can see that the GKP states defined in Eq.~\eqref{eq:physicalGKP} are physical and have finite energy. In the first example, we assume $\ket{\bar{\xi}} = \ket{\bar{0}}$, then
\begin{equation}\label{eq:GKP-0}
\ket{\tilde{0}} = N_0 \int {\rm d} u {\rm d} v \, \eta (u,v) e^{ - i u \hat{p} + i v \hat{q} } \ket{\bar{0}}.
\end{equation}
Consider the wave function of state $\ket{\tilde{0}}$ in position space,
\begin{eqnarray}
\tilde{\psi}_0 (q) &=& \langle q \ket{\tilde{0}} = N_0 \int {\rm d} u {\rm d} v \, \eta (u,v)  \bra{q} e^{ - i u \hat{p} + i v \hat{q} } \ket{\bar{0}} \nonumber = N_0 \sum_{n = - \infty}^{+ \infty} \int {\rm d} u {\rm d} v \, \eta (u,v) e^{- i u v/2} e^{i v q} \langle q  \ket{ 2 n \sqrt{\pi} + u}_q 
\nonumber\\
&=& 
N_0 \sum_{n = - \infty}^{+ \infty} \int {\rm d} v \, \eta (q - 2 n \sqrt{\pi}, v) e^{- i (q - 2 n \sqrt{\pi}) v/2} e^{i v q} \nonumber = \sqrt{2} N_0 \sqrt{\frac{\kappa}{\Delta}} \sum_{n = - \infty}^{+ \infty} e^{-\frac{(q - 2 n \sqrt{\pi})^2}{2\Delta^2}} e^{-\frac{ \kappa^2 (q + 2 n \sqrt{\pi})^2}{8}} \nonumber\\
&=& \sqrt{2} N_0 \sqrt{\frac{\kappa}{\Delta}} \sum_{n = - \infty}^{+ \infty} \exp \left \{-\frac{4\pi n^2 \kappa^2}{2(1+\Delta^2 \kappa^2/4)} \right \} \exp \left \{- \frac{1+\Delta^2 \kappa^2/4}{2 \Delta^2} \left[q - \left( \frac{1-\Delta^2 \kappa^2/4}{1+\Delta^2 \kappa^2/4} \right) 2 n \sqrt{\pi} \right]^2 \right \}. 
\end{eqnarray}
It is evident that the wave function $\tilde{\psi}_0 (q)$ is a sum of a sequence of Gaussian functions weighted by a function that rapidly decreases when $n$ increases, therefore the wave function is normalizable and is physical. Note that the spacing between the Gaussian peaks is slightly modified, 
\begin{eqnarray}
2 \sqrt{\pi} \rightarrow  2 \sqrt{\pi} \left( \frac{1-\Delta^2 \kappa^2/4}{1+\Delta^2 \kappa^2/4} \right), 
\end{eqnarray}
and the variance is also slightly changed,
\begin{eqnarray}
\Delta^2 \rightarrow \frac{\Delta^2}{1+\Delta^2 \kappa^2/4}. 
\end{eqnarray}
When both $\Delta$ and $\kappa$ are sufficiently small, the higher order term $\Delta^2 \kappa^2$ can be neglected, then the wave function $\tilde{\psi}_0 (q)$ can be approximated as
\begin{eqnarray}\label{eq:GKP-0-P-app}
\tilde{\psi}_0 (q) \approx
\sqrt{2} N_0 \sqrt{\frac{\kappa}{\Delta}} \sum_{n = - \infty}^{+ \infty} e^{ - 2 \pi n^2 \kappa^2} e^{-\left( q - 2 n \sqrt{\pi} \right )^2/ 2 \Delta^2} \nonumber\approx \left( \frac{4 \kappa^2}{\pi \Delta^2}\right)^{1/4} \sum_{n = - \infty}^{+ \infty} e^{ - 2 \pi n^2 \kappa^2} e^{-\left( q - 2 n \sqrt{\pi} \right )^2/ 2 \Delta^2},
\end{eqnarray}
where we have used the approximation that $N_0^2 \approx 1/\sqrt{\pi}$ when  $\Delta$ and $\kappa$ are small. 

Then the wave function of state $\ket{\tilde{0}}$ in the momentum space can be calculated in a similar way, 
\begin{eqnarray}
\tilde{\psi}_0 (p) &=& \langle p \ket{\tilde{0}} = N_0 \int {\rm d} u {\rm d} v \, \eta (u,v)  \bra{p} e^{ - i u \hat{p} + i v \hat{q} } \ket{\bar{0}} \nonumber = N_0 \sum_{n = - \infty}^{+ \infty} \int {\rm d} u {\rm d} v \, \eta (u,v) e^{ i u v/2} e^{- i u p} \langle p  \ket{ n \sqrt{\pi} + v}_p \nonumber\\
&=& 
N_0 \sum_{n = - \infty}^{+ \infty} \int {\rm d} v \, \eta (u, p - n \sqrt{\pi}) e^{i (p - n \sqrt{\pi}) u/2} e^{-i u p} \nonumber = \sqrt{2} N_0 \sqrt{\frac{\Delta}{\kappa}} \sum_{n = - \infty}^{+ \infty} e^{-\frac{(p - n \sqrt{\pi})^2}{2\kappa^2}} e^{-\frac{ \Delta^2 (p + n \sqrt{\pi})^2}{8}} 
\nonumber\\
&=& 
\sqrt{2} N_0 \sqrt{\frac{\Delta}{\kappa}} \sum_{n = - \infty}^{+ \infty} \exp \left \{-\frac{ \pi n^2 \Delta^2}{2(1+\Delta^2 \kappa^2/4)} \right \} \exp \left \{- \frac{1+\Delta^2 \kappa^2/4}{2 \kappa^2} \left[p - \left( \frac{1-\Delta^2 \kappa^2/4}{1+\Delta^2 \kappa^2/4} \right) n \sqrt{\pi} \right]^2 \right \}. 
\end{eqnarray}
Note that the spacing between the Gaussian peaks is slightly modified, 
\begin{eqnarray}
\sqrt{\pi} \rightarrow  \sqrt{\pi} \left( \frac{1-\Delta^2 \kappa^2/4}{1+\Delta^2 \kappa^2/4} \right), 
\end{eqnarray}
and the variance is also slightly changed,
\begin{eqnarray}
\kappa^2 \rightarrow \frac{\kappa^2}{1+\Delta^2 \kappa^2/4}. 
\end{eqnarray}
When both $\Delta$ and $\kappa$ are small, the higher order term $\Delta^2 \kappa^2$ can be neglected, then the wave function $\tilde{\psi}_0 (p)$ can be approximated as
\begin{eqnarray}\label{eq:GKP-0-P-app}
\tilde{\psi}_0 (p) \approx \left( \frac{4 \Delta^2}{\pi \kappa^2}\right)^{1/4} \sum_{n = - \infty}^{+ \infty} e^{ - \pi n^2 \Delta^2/2} e^{-\left( p - n \sqrt{\pi} \right )^2/ 2 \kappa^2}. 
\end{eqnarray}

In the second example, we consider 
$\ket{\bar{\xi}} = \ket{\bar{1}}$, then
\begin{equation}\label{eq:GKP-1}
\ket{\tilde{1}} = N_1 \int {\rm d} u {\rm d} v \, \eta (u,v) e^{ - i u \hat{p} + i v \hat{q} } \ket{\bar{1}}.
\end{equation}
Using the expression of $\ket{\bar{1}} = \sum_{n} \ket{(2n + 1)\sqrt{\pi}}_q$ in position space, we can similarly derive its wave function,
\begin{eqnarray}
&&\tilde{\psi}_1 (q) = \langle q \ket{\tilde{1}} = N_1 \int {\rm d} u {\rm d} v \, \eta (u,v)  \bra{q} e^{ - i u \hat{p} + i v \hat{q} } \ket{\bar{1}} \nonumber= \sqrt{2} N_1 \sqrt{\frac{\kappa}{\Delta}} \sum_{n = - \infty}^{+ \infty} e^{-\frac{ \left[ q - (2n+1) \sqrt{\pi}\, \right]^2}{2\Delta^2}} 
e^{-\frac{ \kappa^2 \left[ q + (2n+1) \sqrt{\pi} \right]^2}{8}} \nonumber\\
&&= \sqrt{2} N_1 \sqrt{\frac{\kappa}{\Delta}} \sum_{n = - \infty}^{+ \infty} \exp \left \{-\frac{\pi (2n+1)^2 \kappa^2}{2(1+\Delta^2 \kappa^2/4)} \right \} \exp \left \{- \frac{1+\Delta^2 \kappa^2/4}{2 \Delta^2} \left[q - \left( \frac{1-\Delta^2 \kappa^2/4}{1+\Delta^2 \kappa^2/4} \right) (2n+1) \sqrt{\pi} \right]^2 \right \}. 
\end{eqnarray}
When both $\Delta$ and $\kappa$ are sufficiently small, the higher order term $\Delta^2 \kappa^2$ can be neglected, then the wave function $\tilde{\psi}_1 (q)$ can be approximated as
\begin{eqnarray}\label{eq:GKP-0-P-app}
\tilde{\psi}_1 (q) \approx
\sqrt{2} N_1 \sqrt{\frac{\kappa}{\Delta}} \sum_{n = - \infty}^{+ \infty} e^{ - \frac{(2n+1)^2 \pi  \kappa^2}{2}} e^{- \frac{\left[ q - (2n+1) \sqrt{\pi} \right ]^2}{2 \Delta^2}} \approx \left( \frac{4 \kappa^2}{\pi \Delta^2}\right)^{1/4} \sum_{n = - \infty}^{+ \infty} e^{ - \frac{(2n+1)^2 \pi  \kappa^2}{2}} e^{- \frac{\left[ q - (2n+1) \sqrt{\pi} \right ]^2}{2 \Delta^2}}, 
\end{eqnarray}
where we have used the approximation that $N_1^2 \approx 1/\sqrt{\pi}$ when  $\Delta$ and $\kappa$ are small. 

The wave function in momentum space can be calculated in a similar way,
\begin{eqnarray}
&&\tilde{\psi}_1 (p) = \langle p \ket{\tilde{0}} = N_1 \int {\rm d} u {\rm d} v \, \eta (u,v)  \bra{p} e^{ - i u \hat{p} + i v \hat{q} } \ket{\bar{1}}= \sqrt{2} N_1 \sqrt{\frac{\Delta}{\kappa}} \sum_{n = - \infty}^{+ \infty} (-1)^n e^{-\frac{(p - n \sqrt{\pi})^2}{2\kappa^2}} e^{-\frac{ \Delta^2 (p + n \sqrt{\pi})^2}{8}} \nonumber\\
&&= \sqrt{2} N_1 \sqrt{\frac{\Delta}{\kappa}} \sum_{n = - \infty}^{+ \infty} (-1)^n \exp \left \{-\frac{ \pi n^2 \Delta^2}{2(1+\Delta^2 \kappa^2/4)} \right \} \exp \left \{- \frac{1+\Delta^2 \kappa^2/4}{2 \kappa^2} \left[p - \left( \frac{1-\Delta^2 \kappa^2/4}{1+\Delta^2 \kappa^2/4} \right) n \sqrt{\pi} \right]^2 \right \}. 
\end{eqnarray}
When both $\Delta$ and $\kappa$ are small, the higher order term $\Delta^2 \kappa^2$ can be neglected, then the wave function $\tilde{\psi}_1 (p)$ can be approximated as
\begin{eqnarray}\label{eq:GKP-0-P-app}
\tilde{\psi}_1 (p) \approx \left( \frac{4 \Delta^2}{\pi \kappa^2}\right)^{1/4} \sum_{n = - \infty}^{+ \infty} (-1)^n e^{ - \pi n^2 \Delta^2/2} e^{-\left( p - n \sqrt{\pi} \right )^2/ 2 \kappa^2}. 
\end{eqnarray}

According to the definition of the Wigner function in Eq.~\eqref{eq:WignerDf} and the expression for $\tilde{\psi}_0(q)$, it is straightforward to calculate the Wigner function of the physical GKP state $\ket{\tilde{0}}$,
\begin{eqnarray}\label{eq:Wigner-0-phy-exact}
W(q, p; \ket{\tilde{0}}\bra{\tilde{0}}) &=& \sqrt{\pi} N_0^2 \sum_{m, n} e^{- \pi \Delta_s^2 m^2/4 - 4 \pi \kappa_s^2 n^2 }
\exp \left \{ - \frac{(p - m \sqrt{\pi} \gamma /2)^2}{\kappa_s^2}  - \frac{(q - 2 n \sqrt{\pi} \gamma)^2}{\Delta_s^2}  \right \} \nonumber\\
&& + \sqrt{\pi} N_0^2 \sum_{m, n} (-1)^m e^{- \pi \Delta_s^2 m^2/4 - \pi \kappa_s^2 (2n+1)^2 }
\exp \left\{ - \frac{(p - m \sqrt{\pi} \gamma /2)^2}{\kappa_s^2}  - \frac{ [ q - (2n+1) \sqrt{\pi} \gamma \, ]^2}{\Delta_s^2}  \right \},
\end{eqnarray}
where $\Delta_s = \Delta/\sqrt{1+\Delta^2 \kappa^2/4}$, $\kappa_s = \kappa/\sqrt{1+\Delta^2 \kappa^2/4}$ and $\gamma = \frac{1-\Delta^2 \kappa^2/4}{1+\Delta^2 \kappa^2/4}$. When both $\Delta$ and $\kappa$ are sufficiently small, the higher order term $\Delta^2 \kappa^2$ can be neglected, namely, $\Delta_s \rightarrow \Delta$, $\kappa_s \rightarrow \kappa$ and $\gamma \rightarrow 1$. 
 The Wigner function of the physical GKP state can be approximated as
\begin{eqnarray}\label{eq:Wigner-0-phy-app}
W(q, p; \ket{\tilde{0}}\bra{\tilde{0}}) &\approx& \sum_{m, n} e^{- \pi \Delta^2 m^2/4 - 4 \pi \kappa^2 n^2 }
\exp \left \{ - \frac{(p - m \sqrt{\pi}/2)^2}{\kappa^2}  - \frac{(q - 2 n \sqrt{\pi})^2}{\Delta^2}  \right \} \nonumber\\
&& + \sum_{m, n} (-1)^m e^{- \pi \Delta^2 m^2/4 - \pi \kappa^2 (2n+1)^2 }
\exp \left\{ - \frac{(p - m \sqrt{\pi}/2)^2}{\kappa^2}  - \frac{ [ q - (2n+1) \sqrt{\pi}\, ]^2}{\Delta^2}  \right \}. 
\end{eqnarray}

Consider the Wigner function of the ideal GKP state $\ket{\bar{0}}$ after going through a GDC. The density matrix is
\begin{eqnarray}
\hat{\rho} &=& \int {\rm d} u {\rm d} v f(u, v) \hat{D}(u, v) \ket{\bar{0}} \bra{\bar{0}} \hat{D}^\dag (u, v) 
=
\sum_{n, m} \int {\rm d} u {\rm d} v f(u, v) e^{2 i \sqrt{\pi} v (n - m) } \ket{2 n \sqrt{\pi} + u}_q \bra{2 m \sqrt{\pi} + u}.
\end{eqnarray}
According to Eq.~\eqref{eq:WignerDf}, the Wigner function can be calculated as 
\begin{eqnarray}\label{eq:Wigner-0-ideal-app}
W(q, p; \hat{\rho}) &=& \frac{1}{4\pi \sqrt{\pi} \Delta \kappa} \sum_{m, n} 
\exp \left \{ - \frac{(p - m \sqrt{\pi}/2)^2}{\delta_p^2}  - \frac{(q - 2 n \sqrt{\pi})^2}{\delta_q^2}  \right \} 
\nonumber\\
&& 
+  \frac{1}{4\pi \sqrt{\pi} \Delta \kappa} \sum_{m, n} (-1)^m \exp \left\{ - \frac{(p - m \sqrt{\pi}/2)^2}{\delta_p^2}  - \frac{ [ q - (2n+1) \sqrt{\pi}\,]^2}{\delta_q^2}  \right \}. 
\end{eqnarray}

After the action of GDC, the new Wigner function $W(q, p)$ is related to the old Wigner function $W_0(q, p)$ via
\begin{eqnarray}
W(q, p) = \int {\rm d} u {\rm d} v f(u, v) W_0(q+u, p+v) 
= \int {\rm d} u {\rm d} v f(u - q, v - p) W_0(u, v),
\end{eqnarray}
Which implies that the new Wigner function is the convolution of the old Wigner function and the noise distribution function $f$. Although the physical GKP state is not Gaussian, their Wigner function can be written as a sum of a sequence of Gaussian functions. Since the convolution of two Gaussian functions gives also a Gaussian function, the Wigner function of a physical GKP state after going through a GDC is still a sum of a sequence of Gaussian functions. Furthermore, the variances of the new Gaussian functions are the sum of the variances of the old Gaussian functions and those of noise distribution function. Suppose $W_0 (q, p) = W(q, p; \ket{\tilde{0}}\bra{\tilde{0}})$, then the Wigner function after going through the GDC is given by
\begin{eqnarray}\label{eq:Wigner-0-GDC}
W_{\rm GDC}(q, p; \ket{\tilde{0}}\bra{\tilde{0}}) &\approx&  \frac{\Delta \kappa}{\sqrt{(\delta_q^2+\Delta^2)(\delta_p^2+\kappa^2)}} \bigg\{ \sum_{m, n} e^{- \pi \Delta^2 m^2/4 - 4 \pi \kappa^2 n^2 }
\exp \left [ - \frac{(p - m \sqrt{\pi}/2)^2}{\kappa^2 + \delta_p^2}  - \frac{(q - 2 n \sqrt{\pi})^2}{\Delta^2+\delta_q^2}  \right ] \nonumber\\
&& + \sum_{m, n} (-1)^m e^{- \pi \Delta^2 m^2/4 - \pi \kappa^2 (2n+1)^2 }
\exp \left[ - \frac{(p - m \sqrt{\pi}/2)^2}{\kappa^2+\delta_p^2}  - \frac{ ( q - (2n+1) \sqrt{\pi}\, )^2}{\Delta^2+\delta_q^2}  \right ] \bigg\}. 
\end{eqnarray}

\section{Error distribution of GKP state after SUM gate with physical ancillary qubit}\label{sec:appB}

We give the detailed calculation of the error distribution of GKP state after GKP error correction with a physical ancillary qubit, i.e., we calculate the probability distribution of the variable $u'$ given by Eq.~\eqref{eq:error distrbution after SUM gate}, with the probability distribution of $u_1$ and $u_2$ given by Eq.~\eqref{eq:distribution u1u2}.
We first need to calculate the probability $P(u'\le x)$ for a given $x$,
\begin{eqnarray}
P(u'\le x) &=& \sum_{k}^{}P\left [ u_2 \ge k\sqrt{\pi}-x \text{ and } (k-\frac{1}{2})\sqrt{\pi}\le u_1+u_2 < (k+\frac{1}{2})\sqrt{\pi}  \right ] 
\nonumber \\
&=&
\sum_{k}^{}\int_{k\sqrt{\pi}-x}^{+\infty }f_{q_2}(u_2){\rm d}u_2\int_{(k-1/2)\sqrt{\pi}-u_2}^{(k+1/2)\sqrt{\pi}-u_2}f_{q_1}(u_1){\rm d}u_1
\nonumber \\
&=&
\frac{1}{2} \sum_{k}^{} \int_{k\sqrt{\pi}-x}^{+\infty }{\rm d}u_2f_{q_2}(u_2) \left [ \text{erf}\left ( \frac{(k+1/2)\sqrt{\pi}-u_2}{\Delta}  \right ) -\text{erf}\left ( \frac{(k-1/2)\sqrt{\pi}-u_2}{\Delta} \right )  \right ].
\end{eqnarray}
Then the probability distribution of $u'$ is obtained by taking derivative with respect to $x$,
\begin{eqnarray}
F(u'=x) &=&
\frac{\mathrm{d} P(u'\le x)}{\mathrm{d} x} 
\nonumber\\ 
&=&
\frac{1}{2}\sum_{k}^{}f_{q_2}(u_2=k\sqrt{\pi}-x)\cdot \left [ \text{erf}\left ( \frac{\sqrt{\pi}/2+x}{\Delta}  \right ) -\text{erf}\left ( \frac{-\sqrt{\pi}/2+x}{\Delta} \right )  \right ]  
\nonumber\\
&=&
\frac{1}{2\sqrt{\pi}\tilde {\Delta} }  \left [ \text{erf}\left ( \frac{\sqrt{\pi}/2+x}{\Delta}  \right ) -\text{erf}\left ( \frac{-\sqrt{\pi}/2+x}{\Delta} \right )  \right ] \cdot \sum_{t}^{}\text{exp}\left [ -\frac{(x -t\sqrt{\pi})^2}{\tilde{\Delta}^2} \right ]. 
\end{eqnarray}
Finally Eq.~\eqref{eq:udot distribution} is obtained by simply rewriting the above result.

\section{Calculation of the failure probability of $n$-qubit GKP repetition code}\label{sec:appC}

The correspondence between measurement outcomes and correctable errors is given by Tab.~\ref{tab:GKP nrep code}. However, this decoding procedure may result in misidentification of the error, which is different from that of the classical $n$-qubit repetition code. Here we provide the detailed calculation of the failure probability of $n$-qubit GKP repetition code. Similar to the discussion in Sec.~\ref{sec:4A}, we need to reverse the decoding process and impose some conditions to be satisfied. All possible cases are summarized as follows:
\begin{itemize}
\item Case 1: If no error occurs $\Rightarrow$we require $M_1,M_2, \cdots,M_{n-1}\in \text{NPZ}$; 
\item Case 2: If $\bar{X}$ applies on data qubit $D_1$ $\Rightarrow$ we require $M_1,M_2, \cdots,M_{n-1}\in \text{PZ}$. We find that this failure probability is the same as all $C_{n}^1$ cases where $\bar{X}$ applies on a single data qubit.
\item Case $i$ ($3\le i \le \frac{n+1}{2}$): If $\bar{X}$ applies on data qubit $D_1,D_2, \cdots, D_{i-1}$ $\Rightarrow$ we require $M_1, \cdots, M_{i-2}\in \text{NPZ}, M_{i-1}, \cdots, M_{n-1}\in \text{PZ}$. This failure probability is the same as all $C_{n}^{i-1}$ cases where $\bar{X}$ applies on $i-1$ data qubits.
\item Case $\frac{n+3}{2}$: If errors occur on more than $(n-1)/2$ data qubits, with probability $\sum_{j=\frac{n+1}{2} }^{n} C_n^j P_F^j (1-P_F)^{n-j}$ $\Rightarrow$ the error correction fails. 
\end{itemize}
Note that we incorporate all $C_{n}^{s}$ possibilities where errors occur on $s$ data qubits into one case, and we consider a representative where errors occur on the first $s$ data qubits $D_1,D_2, \cdots, D_s$, for all possibilities have the same failure probability.

Now we calculate the failure probability for these $(n+3)/2$ cases, the sum of which gives the total probability of failure. Consider case 1, there are $2n-1$ constraints needed to be satisfied simultaneously,
\begin{eqnarray}\label{eq:NoErrorDomain nrep}
\text{No Pauli $\bar{X}$ error} &\Rightarrow& \left | u_1'-2s_1\sqrt{\pi}  \right |<\frac{\sqrt{\pi}}{2}, \left | u_2'-2s_2\sqrt{\pi}  \right |<\frac{\sqrt{\pi}}{2}, \cdots, \left | u_n'-2s_n\sqrt{\pi}  \right |<\frac{\sqrt{\pi}}{2}, \nonumber\\
M_1,M_2, \cdots, M_{n-1}\in \text{NPZ} &\Rightarrow& \left | u_1'+u_2'+\alpha _1-2t_1\sqrt{\pi}  \right |<\frac{\sqrt{\pi}}{2}, \cdots, \left | u_1'+u_n'+\alpha_{n-1}-2t_{n-1}\sqrt{\pi}  \right |<\frac{\sqrt{\pi}}{2},
\end{eqnarray}
where $s_i \in \mathbb{Z}$ and $t_i \in \mathbb{Z}$. The probability of success is obtained by integrating the probability distribution of $2n-1$ variables in the domain defined by these $2n-1$ inequalities. Similar to the discussion in Sec.~\ref{sec:4A}, we use the numerical method to calculate the integration. We first fix a point $(u_1',u_2',...,u_n')$ defined by the first $n$ inequalities in Eq.~\eqref{eq:NoErrorDomain nrep}, then success probability at this given point is
\begin{eqnarray}
P_\alpha ^1(u_1', \cdots, u_n') 
&=&
\left ( \sum_{t_1}^{}\int_{-\sqrt{\pi}/2+2t_1\sqrt{\pi}-u_1'-u_2'}^{\sqrt{\pi}/2+2t_1\sqrt{\pi}-u_1'-u_2'}f_{q_1'}(\alpha _1) {\rm d} \alpha _1   \right )\times \cdots \times \left ( \sum_{t_{n-1}}^{}\int_{-\sqrt{\pi}/2+2t_{n-1}\sqrt{\pi}-u_1'-u_n'}^{\sqrt{\pi}/2+2t_{n-1}\sqrt{\pi}-u_1'-u_n'}f_{q_{n-1}'}(\alpha_{n-1}) {\rm d} \alpha_{n-1} \right )
\nonumber\\
&\approx& 
\frac{1}{2^{n-1}}\prod_{k=2}^{n}\left [ {\rm erf} \left ( \frac{\sqrt{\pi}/2-u_1'-u_k' }{\tilde {\Delta}}  \right )-{\rm erf} \left ( \frac{-\sqrt{\pi}/2-u_1'-u_k' }{\tilde {\Delta}} \right )   \right ],
\end{eqnarray}
where we have only kept one term $t_1=t_2=\cdots=t_{n-1}=0$ in the summation because the contribution from other terms is negligible. Then the failure probability of case 1 is given by integrating the failure probability $1-P_\alpha ^1(u_1',...,u_n')$ over all points satisfying the constraints in Eq.~\eqref{eq:NoErrorDomain nrep}, weighted by the probability distribution $\prod_{i=1}^{n}F(u_i')$,
\begin{eqnarray}
P_{f,\text{n-rep}}^1 &=&
\int_{u_1'\in {\rm NPZ}}^{} \cdots \int_{u_n'\in {\rm NPZ}}^{} \left [\prod_{i=1}^{n}F(u_i')  \right ]  \left [ 1-P_{\alpha }^1(u_1', \cdots, u_n') \right ] {\rm d} u_1' ... {\rm d} u_n' 
\nonumber\\ 
&\approx& 
\int_{u_1'=-\sqrt{\pi}/2 }^{\sqrt{\pi}/2 } \cdots \int_{u_n'=-\sqrt{\pi}/2 }^{\sqrt{\pi}/2 } \left [\prod_{i=1}^{n}F(u_i')  \right ]\left [ 1-P_{\alpha }^1(u_1', \cdots, u_n') \right ] {\rm d} u_1' ...{\rm d} u_n', 
\end{eqnarray}
where we have only kept one term with $s_1=s_2=\cdots=s_n=0$ in the summation because the contribution form other terms is negligible.

Similarly, we can derive the failure probability of case 2 by taking into account the condition that $u_1'\in \text{PZ},u_2'\in \text{NPZ}, \cdots, u_n'\in \text{NPZ}$,
\begin{eqnarray}
P_{f,\text{n-rep}}^2 &=&
\int_{u_1'\in {\rm PZ}}^{}\int_{u_2'\in {\rm NPZ}}^{} \cdots \int_{u_n'\in {\rm NPZ}}^{} \left [\prod_{i=1}^{n}F(u_i')  \right ]  \left [ 1-P_{\alpha }^2(u_1', \cdots, u_n') \right ] {\rm d} u_1' \cdots {\rm d} u_n' 
\nonumber\\ 
&\approx& 
2\int_{u_1'=\sqrt{\pi}/2 }^{3\sqrt{\pi}/2 }\int_{u_2'=-\sqrt{\pi}/2 }^{\sqrt{\pi}/2 } \cdots \int_{u_n'=-\sqrt{\pi}/2 }^{\sqrt{\pi}/2 } \left [\prod_{i=1}^{n}F(u_i')  \right ]\left [ 1-P_{\alpha }^2(u_1', \cdots, u_n') \right ] {\rm d} u_1' \cdots {\rm d} u_n', 
\end{eqnarray}
where $P_{\alpha }^2(u_1', \cdots, u_n')$ is the success probability for a given point $(u_1',u_2', \cdots, u_n')$ when $M_1,M_2, \cdots, M_{n-1}\in {\rm PZ}$,
\begin{eqnarray}
P_{\alpha }^2(u_1', \cdots, u_n') 
&=& \left ( \sum_{t_1}^{}\int_{\sqrt{\pi}/2+2t_1\sqrt{\pi}-u_1'-u_2'}^{3\sqrt{\pi}/2+2t_1\sqrt{\pi}-u_1'-u_2'}f_{q_1'}(\alpha _1) {\rm d} \alpha _1   \right )\times \cdots \times \left ( \sum_{t_{n-1}}^{}\int_{\sqrt{\pi}/2+2t_{n-1}\sqrt{\pi}-u_1'-u_n'}^{3\sqrt{\pi}/2+2t_{n-1}\sqrt{\pi}-u_1'-u_n'}f_{q_{n-1}'}(\alpha_{n-1}) {\rm d} \alpha_{n-1} \right )
\nonumber\\
&\approx& 
\frac{1}{2^{n-1}}\prod_{k=2}^{n}\left [ {\rm erf} \left ( \frac{3\sqrt{\pi}/2-u_1'-u_k' }{\tilde {\Delta}}  \right )-{\rm erf} \left ( \frac{\sqrt{\pi}/2-u_1'-u_k' }{\tilde {\Delta}} \right )   \right ].
\end{eqnarray}
Note that there are $C_n^1$ cases giving the same result as case 2, so we need to plus $P_{f,\text{n-rep}}^2$ with $C_n^1$ in the total failure probability.

In a similar way, the failure probability of case $i$ ($3\le i\le \frac{n+1}{2}$) is given by taking into account the condition that $u_1', \cdots, u_{i-1}'\in {\rm PZ}$, $u_i', \cdots, u_n'\in {\rm NPZ}$,
\begin{eqnarray}
P_{f,\text{n-rep}}^i &=& \int_{u_1'\in {\rm PZ}}^{}\cdots\int_{u_{i-1}'\in {\rm PZ}}^{}\int_{u_i'\in {\rm NPZ}}^{} \cdots \int_{u_n'\in {\rm NPZ}}^{} \left [\prod_{i=1}^{n}F(u_i')  \right ]  \left [ 1-P_{\alpha }^i(u_1', \cdots, u_n') \right ] {\rm d} u_1' \cdots {\rm d} u_n' 
\nonumber \\
&\approx& 2^{i-1}\int_{u_1'=\sqrt{\pi}/2 }^{3\sqrt{\pi}/2 }\cdots\int_{u_{i-1}'=\sqrt{\pi}/2 }^{3\sqrt{\pi}/2 }\int_{u_i'=-\sqrt{\pi}/2 }^{\sqrt{\pi}/2 } \cdots \int_{u_n'=-\sqrt{\pi}/2 }^{\sqrt{\pi}/2 } \left [\prod_{i=1}^{n}F(u_i')  \right ]\left [ 1-P_{\alpha }^i(u_1', \cdots, u_n') \right ] {\rm d} u_1' \cdots {\rm d} u_n',
\nonumber \\
\end{eqnarray}
where $P_{\alpha }^i(u_1', \cdots, u_n')$ is the success probability for the point $(u_1',u_2', \cdots, u_n')$ when $M_1, \cdots, M_{i-2}\in \text{NPZ}$, $M_{i-1}, \cdots, M_{n-1}\in \text{PZ}$,
\begin{eqnarray}
P_{\alpha}^i(u_1', \cdots, u_n') &=& \frac{1}{2^{n-1}}\prod_{k_1=2}^{i-1}\left [ {\rm erf} \left ( \frac{5\sqrt{\pi}/2-u_1'-u_{k_1}' }{\tilde {\Delta}}  \right )-{\rm erf} \left ( \frac{3\sqrt{\pi}/2-u_1'-u_{k_1}'}{\tilde {\Delta}} \right )   \right ]
\nonumber \\
&=&
\times \prod_{k_2=i}^{n}\left [ {\rm erf} \left ( \frac{3\sqrt{\pi}/2-u_1'-u_{k_2}' }{\tilde {\Delta}}  \right )-{\rm erf} \left ( \frac{\sqrt{\pi}/2-u_1'-u_{k_2}' }{\tilde {\Delta}} \right )   \right ].  
\end{eqnarray}
There are $C_n^{i-1}$ cases giving the same result as the case $i$, so we need to add a factor $C_n^{i-1}$ in the expression of the failure probability.

The failure probability of case $\frac{n+3}{2}$ is given by
\begin{equation}
P_{f,\text{n-rep}}^{\frac{n+3}{2}}=\sum_{j=\frac{n+1}{2} }^{n} C_n^j P_F^j (1-P_F)^{n-j}.
\end{equation}

Finally, the total failure probability of the $n$-qubit GKP repetition code is summation of the failure probabilities of all $(n+3)/2$ cases,
\begin{equation}
P_{f, \text{n-rep}}=P_{f, \text{n-rep}}^1+C_n^1 P_{f, \text{n-rep}}^2+\sum_{i=3}^{\frac{n+1}{2}} C_n^{i-1} P_{f, \text{n-rep}}^i+P_{f,\text{n-rep}}^{\frac{n+3}{2}}. 
\end{equation}

\end{widetext} 

\bibliography{ref_GKP-repetition}

\end{document}